\documentclass[11pt,a4paper]{article}
\usepackage{amssymb}
\usepackage{amsmath}
\usepackage{graphics}
\usepackage{epsfig}
\usepackage{float}
\usepackage{graphicx}
\usepackage{amsfonts}
\usepackage{amsthm}
\usepackage{newlfont}
\usepackage{color}
\usepackage[numbers]{natbib}

\begin{document}

\title{Interactions of $B_{c}$ Meson in Relativistic Heavy-Ion Collisions}
\author{Shaheen Irfan$^{a,b}$\thanks{%
shaheen.irfan@lums.edu.pk },
Faisal Akram$^{b}$\thanks{%
faisal.chep@pu.edu.pk} and Bilal Masud$^{b}$\thanks{%
bilalmasud.chep@pu.edu.pk}\\
$^{a}$\textit {Department of Physics, Syed Babr Ali School of Science $\&$ Engineering,}\\
\textit{LUMS, D.H.A Lahore 54792, Pakistan.} \\
$^{b}$\textit {Center for High Energy Physics,  Punjab University, Lahore,$54590$, Pakistan.}}
\maketitle

\begin{abstract}
We calculate the dissociation cross-sections of $B_{c}$ mesons by $\pi$ and $\rho$ mesons including anomalous processes using an effective hadronic Lagrangian. The enhancement of $B_c$ production is expected due to QGP formation in heavy-ion experiments. However it is also expected that the production rate of $B_c$ meson can be affected due to the interaction with comovers. These processes are relevant for the experiments at RHIC. Thermal average cross-sections of $B_{c}$ are evaluated with a form factor when a cut off parameter in it is 1 and 2 GeV. Using these thermal average cross-sections in the kinetic equation we study the time evolution of $B_{c}$ mesons due to dissociation in the hadronic matter formed at RHIC.
\end{abstract}

\bigskip

\noindent PACS number(s): 25.75.-q, 13.75.Lb, 14.40.Nd

\section{\protect\bigskip \textbf{Introduction}}
In 1986 Matsui and Satz \cite{matsui1986} hypothesized that in a deconfined medium color screening would have dissociated the $J/\psi $, resulting in a suppressed yield of $J/\psi $. This deconfined state is called Quark-Gluon Plasma (QGP). Thus for the existence of QGP, suppression of $J/\psi$ could be considered as a probe. Anomalously large suppression of events was observed by NA50 experiment at CERN \cite{NA50} with moderate to large transfer energy from the Pb+Pb collision at $P_{lab}=158$ GeV/c. However, this observed suppression may also occur due to absorption by comoving hadrons mainly $\pi$ and $\rho$, especially if the dissociation cross section is at least few mb \cite{cassing1997,armesto1998,kahana1999,gale1999,spieles1999,sa1999}. To calculate these cross sections, quark potential models, perturbative QCD \cite{kharzeev1994}, QCD sum-rule approach \cite{sum-rule,quark models} and flavor symmetric effective Lagrangian \cite{lin2000,lin2001,haglin2000,liu2001} has been used.
\noindent Analogous to charmonium, suppression of bottomonium states is also predicted during the formation of QGP \cite{matsui1986}. Recently it was observed by CMS in Pb+Pb collisions that excited states of bottomonium are strongly suppressed \cite{cms}. To have unambiguous interpretation of the the observed signal, the information of dissociation cross section is also needed  \cite {lin2001,vogt1997}. It was suggested that the production rate of heavy mixed flavor hadrons would be affected in the presence of QGP \cite{schro2000,refelski2002}. For  calculating the rate of production of these hadrons comprehensive information is required to distinguish QGP affected hadron production and suppression due to dissociation by comovers. It is expected that $B_{c}$ production could be enhanced in the presence of QGP \cite{refelski2002,lodhi2007}. QGP contains many unpaired $b(\overline{b})$ and $c(\overline{c})$  quarks due to color Debye screening. These unpaired $b(\overline{b})$ and $c(\overline{c})$ quarks upon encounter could form $B_{c}$ $(\overline{b}c)$ or $(b \overline{c})$  mesons and due to relatively large binding energy, $B_{c}$ mesons probably survive in QGP \cite{lodhi2007}. However, observed production rate would also depend upon the dissociation cross section by hadronic comovers.\\

\noindent In Ref. \cite{lodhi2007} $B_{c}$ absorption by nucleons was examined with the meson-baryon exchange model. The calculated cross sections were in the range of a few millibarn. Recently in Ref. \cite{lodhi2011}, using the same couplings and hadronic Lagrangian within meson exchange model the dissociation of $B_c$ meson by $\pi$ meson were examined. The range of the resulting cross sections involving the form factors were $2-7$ mb and $0.2-2$ mb for the processes $B_c^+ +\pi\rightarrow D+B $ and $B_c^+ +\pi\rightarrow D^*+B^*$, respectively. In Ref. \cite{faisal2011}, the dissociation of $B_c$ meson by $\rho$ mesons were examined. For the processes $B_{c}^+ +\rho\rightarrow D^* +B$ and $B_{c}^+ +\rho\rightarrow D+B^*$ the resultant cross sections with the form factor were in the range of $0.6-3$ and $0.05-0.3$ mb, respectively .\\
 In this paper we investigate the $B_{c}$ dissociation by $\pi $\ and $\rho $ mesons including anomalous couplings like PVV, PPPV and PVVV which were ignored in the previous studies. Inclusion of these couplings results in opening of new dissociation channels and addition of new processes and extra diagrams. The contribution of anomalous couplings is found to be significant for calculating cross sections of charmonium dissociation with $\pi$ and $\rho$ meson in Ref. \cite{oh2001}, $K$ mesons in Ref. \cite{azvedo2004} and  dissociation of $B_{c}$ meson by nucleons in Ref. \cite{faisal2012}. We also calculate the thermal average cross sections and  study the time evolution of $B_{c}$ meson at RHIC using a schematic expanding fireball model with an initial $B_c$ abundance determined by the statistical model.
\noindent The paper is organized as follows. In Sec. 2, the interaction Lagrangian terms which are relevant for the description of the dissociation of $B_{c}$ by $\pi$ and $\rho$ mesons including anomalous processes are given and also analytical expressions of the amplitudes for the dissociation of $B_{c}$ meson are reported.  In Sec. 3, we calculate the cross sections with and without form factor and thermal average cross sections. In Sec. 4, we study time evolution of the $B_{c}$ meson abundance at RHIC in a schematic model.  In Sec. 5, we present the summary and discussion.   \bigskip\

\section{\textbf{Interaction Lagrangian} and \textbf{Amplitudes } of \textbf{$B_{c}$ meson dissociation}}
\subsection{Interaction Lagrangian}

We consider the following reactions using an effective Hadronic Lagrangian.
\begin{equation}
\begin{tabular}{llll}
$B_{c}^{+}+\pi\rightarrow  D+B$,         & $B_{c}^{-}+\pi \rightarrow \bar{D} +\bar{B}$ & $ %
B_{c}^{+}+\rho \rightarrow D+B$,          & $B_{c}^{-}+\rho \rightarrow \bar{D}+\bar{B}$ \\
$B_{c}^{+}+\pi \rightarrow D^{\ast}+B,$    & $B_{c}^{-}+\pi \rightarrow \bar{D}^{\ast}+\bar{B},$ & $ %
B_{c}^{+}+\rho \rightarrow D^{\ast}+B,$   & $B_{c}^{-}+\rho \rightarrow \bar{D}^{\ast}+\bar{B},$  \\
$B_{c}^{+}+\pi \rightarrow D+B^{^{\ast}}$, & $B_{c}^{-}+\pi \rightarrow \bar{D}+ \bar{B}^{\ast},$ & $ %
B_{c}^{+}+\rho \rightarrow D+B^{\ast},$  & $B_{c}^{-}+\rho \rightarrow \bar{D}+\bar{B}^{\ast},$  \\
$B_{c}^{+}+\pi \rightarrow D^{\ast}+B^{\ast},$ & $B_{c}^{-}+\pi \rightarrow \bar{D}^{\ast}+\bar{B}^{\ast},$ & $ %
B_{c}^{+}+\rho \rightarrow D^{\ast}+B^{\ast},$ & $B_{c}^{-}+\rho \rightarrow \bar{D}^{\ast}+\bar{B}^{\ast}.$ %
\end{tabular}
\label{1}
\end{equation}
\\
\noindent The processes in the first and second column, and also  in the third and fourth column have same cross sections as being charge conjugation of each other. The generic form for the 1st reaction is given as
\begin{equation}
\begin{tabular}{llll}
$B_{c}^{+}+\pi^{+}\rightarrow  D^{+}+B^{+}$, $B_{c}^{+}+\pi^{-}\rightarrow  D^{0}+B^{0}$, $B_{c}^{+}+\pi^{0}\rightarrow  D^{+}+B^{0}$ &  $B_{c}^{+}+\pi^{0}\rightarrow  D^{0}+B^{+}$
\end{tabular}
\label{1}
\end{equation}

\noindent For calculating the cross sections of the above reactions, relevant interaction Lagrangian terms are required. The required interaction Lagrangian for normal processes (for which the relevant couplings are dimensionless) are obtained using the method described in Refs. \cite{lodhi2011, faisal2011} and are given as follows.

\begin{subequations}
\label{eq2}
\begin{eqnarray}
\mathcal{L}_{\pi DD^{\ast }} &=&ig_{\pi DD^{\ast }}\;D^{\ast \mu } \vec{\tau} %
\cdot (\bar{D}\partial _{\mu }\vec{\pi} -\partial _{\mu }%
\bar{D}\vec{\pi} )+hc  \label{3.1a} \\
\mathcal{L}_{\pi BB^{\ast }} &=&ig_{\pi BB^{\ast }}\;\bar{B}^{\ast \mu }%
\vec{\tau} \cdot (B\partial _{\mu }\vec{\pi}-\partial
_{\mu }B \vec{\pi} )+hc  \label{2b} \\
\mathcal{L}_{B_{c}BD^{\ast }} &=& ig_{B_{c}BD^{\ast }}\;\bar{D}^{\ast\mu}\;(B_{c}^{-}\partial
_{\mu}\bar{B}-\partial _{\mu }B_{c}^{-}\bar{B})+hc  \label{2c} \\
\mathcal{L}_{B_{c}B^{\ast }D} &=&ig_{B_{c}B^{\ast }D}\;B^{\ast \mu}\;
(B_{c}^{-}\partial_{\mu}D-\partial _{\mu}B_{c}^{-}D%
)+hc  \label{2d} \\
\mathcal{L}_{\pi B_{c}D^{\ast }B^{\ast }} &=&-g_{\pi B_{c}D^{\ast }B^{\ast }}B_{c}^{+}%
\bar{B}^{\ast \mu }\vec{\tau} \cdot \vec{\pi} %
\bar{D}_{\mu }^{\ast }+hc  \label{2e} \\
\mathcal{L}_{\rho DD} &=&ig_{\rho DD}(D  \vec{\tau}\partial _{\mu }%
\bar{D}-\partial _{\mu }D \vec{\tau}  \bar{D})\cdot
\vec{\rho} ^{\mu },  \label{2f} \\
\mathcal{L}_{\rho BB} &=&ig_{\rho BB} \ \  \vec{\rho}^{\mu }\cdot(\bar{B}\vec{\tau} %
\partial _{\mu }B-\partial _{\mu }\bar{B}\vec{\tau} B)  \label{2g} \\
\mathcal{L}_{\rho D^{\ast }D^{\ast }} &=&ig_{\rho D^{\ast }D^{\ast }} \ [\vec{\rho}^{\mu }\cdot \left (
\partial _{\mu }D^{\ast \nu }\vec{\tau} \bar{D}_{\nu }^{\ast}
-D^{\ast \nu }\vec{\tau} \partial _{\mu }\bar{D}_{\nu }^{\ast }\right)  \label{2h}\\
&&+  \bar{D}^{\ast \mu } \cdot \left( D^{\ast \nu }\vec{\tau} \cdot \partial _{\mu }\vec{\tau} _{\nu%
}-\partial _{\mu }D^{\ast \nu }\vec{\tau} \cdot \vec{\rho }_{\nu }\right)  \notag \\
&&+D^{\ast \mu }\cdot \left( \vec{\tau} \cdot\vec{\tau} %
^{\nu }\partial _{\mu }\bar{D}_{\nu }^{\ast }-\vec{\tau} %
\cdot \partial _{\mu }\vec{\rho}^{\nu }\bar{D}_{\nu }^{\ast}\right) ] \notag \\
\mathcal{L}_{\rho B^{\ast }B^{\ast }} &=&ig_{\rho B^{\ast }B^{\ast }}[  \vec{\rho}^{\mu } \cdot\left(
\partial _{\mu }\bar{B}^{\ast \nu }\vec{\tau} B_{\nu }^{\ast
}-\bar{B}^{\ast \nu }\vec{\tau} \partial _{\mu }B_{\nu
}^{\ast }\right)  \label{2i} \\
&&+ B^{\ast \mu} \cdot \left( \bar {B}^{\ast \nu }\vec{\tau} \cdot \partial _{\mu }%
\vec{\rho}_{\nu }-\partial _{\mu }\bar{B}^{\ast \nu } %
\vec{\tau} \cdot\vec{\rho}_{\nu }\right)\notag \\
&&+\bar{B}^{\ast \mu }  \cdot\left( \vec{\tau} \cdot
\vec{\rho}^{\nu }\partial _{\mu }B_{\nu }^{\ast }-%
\vec{\tau} \cdot \partial _{\mu }\vec{\rho}^{\nu
}B_{\nu }^{\ast }\right) ] \notag \\
\mathcal{L}_{\rho B_{c}D^{\ast }B} &=&g_{\rho B_{c}D^{\ast }B}\;B_{c}^{+}%
\bar{B}\vec{\tau} \cdot \vec{\rho}_{\mu }%
\bar{D}^{\ast \mu }+hc  \label{2j} \\
\mathcal{L}_{\rho B_{c}DB^{\ast }} &=&g_{\rho B_{c}DB^{\ast }}\;B_{c}^{+}%
\bar{B}^{\ast \mu }\vec{\tau} \cdot \vec{\rho}%
_{\mu }\bar{D}+hc  \label{2k}
\end{eqnarray}%
\end{subequations}

\noindent In addition to the above normal terms there are anomalous terms as well which are required to give a complete description of the hadronic processes. The required interaction Lagrangian for the anomalous processes (for which the relevant couplings are not dimensionless) are obtained using the method described in Ref. \cite{oh2001} and are given as follows.

\begin{subequations}
\label{eq3}
\begin{eqnarray}
\mathcal{L}_{\pi D^{\ast }D^{\ast }} &=&-g_{_{\pi D^{\ast }D^{\ast }}}\varepsilon
^{\mu \nu \alpha \beta }\left[ (\partial _{\mu }{D_{\nu }^{\ast })}\vec{\tau}  \cdot \vec{\pi}
(\partial _{\alpha }\bar{D}_{\beta }^{\ast })\right]   \label{2a} \\
\mathcal{L}_{\pi B^{\ast }B^{\ast }} &=&g_{_{\pi B^{\ast }B^{\ast }}}\varepsilon ^{\mu
\nu \alpha \beta }[\left( \partial _{\alpha }{\bar{B}_{\beta }^{\ast }}%
\right) \vec{\tau}  \cdot \vec{\pi} (\partial _{\mu }B^{\ast \nu })]  \label{2b} \\
\mathcal{L}_{B_{c}D^{\ast }B^{\ast }} &=&g_{_{{B_{c}D^{\ast }B}^{\ast }}}\varepsilon
^{\mu \nu \alpha \beta }[(\partial _{\mu }{D_{\nu }^{\ast }})(\partial
_{\alpha }B^{\ast \beta }){B_{c}}^{-}+{B_{c}}^{+}(\partial _{\alpha }%
\bar{B}^{\ast \beta })(\partial _{\mu }\bar{D}^{\ast \nu })]
\label{2c} \\
\mathcal{L}_{\rho D^{\ast }D} &=&-g_{_{\rho D^{\ast }D}}{\Large \varepsilon }^{\mu \nu
\alpha \beta }\left( D\partial _{\mu }{\Large \rho }{_{\nu }}\partial
_{\alpha }\bar{D}_{\beta }^{\ast }+\partial _{\mu }D_{\nu }^{\ast
}\partial _{\alpha }{\Large \rho }{_{\beta }}\bar{D}\right)   \label{2d}\\
\mathcal{L}_{_{\rho B^{\ast }B}} &=&-g_{_{\rho B^{\ast }B}}{\ }{\Large \varepsilon }^{\mu
\nu \alpha \beta }(B\partial _{\mu }{\Large \rho }{_{\nu }}\partial _{\alpha
}{\large \bar{B}}_{\beta }^{\ast }+\partial _{\mu }B_{\nu }^{\ast }\partial
_{\alpha }{\Large \rho }_{\beta }\bar{B})  \label{2e} \\\;
\mathcal{L}_{\pi B_{c}D^{\ast}B} &=&-ig_{_{\pi B_{c}D^{\ast }B}}{\Large \varepsilon }%
^{\mu \nu \alpha \beta }[D_{\mu }^{\ast }(\partial _{\nu
}B_{c}^{-})(\vec{\tau}  \cdot \partial _{\alpha }\vec{\pi} )(\partial _{\beta }B)+\bar{D_{\mu
}^{\ast }}(\vec{\tau}  \cdot \partial _{\nu }\vec{\pi})(\partial _{\alpha }B_{c}^{+})(\partial
_{\beta }\bar{B})] \label{2f} \\
\mathcal{L}_{\pi B_{c}DB^{\ast }} &=&-ig_{_{\pi B_{c}DB^{\ast }}}{\Large \varepsilon }%
^{\mu \nu \alpha \beta }[B_{\mu }^{\ast }(\partial _{\nu
}B_{c}^{-})(\vec{\tau}  \cdot \partial _{\alpha }\vec{\pi} )(\partial _{\beta }D)+\bar{B_{\mu
}^{\ast }}(\partial _{\nu }B_{c}^{+})(\vec{\tau} \cdot \partial _{\alpha }\vec{\pi}  )(\partial
_{\beta }\bar{D})]  \label{2g} \\
\mathcal{L}_{\rho B_{c}BD} &=&-ig_{_{\rho B_{c}BD}}{\Large \varepsilon }^{\mu \nu
\alpha \beta }[{\large \rho }_{\mu }(\partial _{\nu }D)(\partial _{\alpha
}B)(\partial _{\beta }B_{c}^{-})+{\large \rho }_{\mu }(\partial _{\nu }%
\overline{B})(\partial _{\alpha }\bar{D})(\partial _{\beta }B_{c}^{+})]\label{2h} \\
\mathcal{L}_{\rho B_{c}B^{\ast }D^{\ast}} &=&ig_{\rho B_{c}{D^{\ast }B}^{\ast}}%
{\Large \varepsilon }^{\mu \nu \alpha \beta }[B_{\mu }^{\ast }{\Large \rho }{%
_{\nu }}D_{\alpha }^{\ast }(\partial _{\beta }B_{c}^{-})+\bar{D}_{\mu
}^{\ast }{\Large \rho }{_{\nu }}{\large \bar{B}}_{\alpha }^{\ast }(\partial
_{\beta }B_{c}^{+})]  \label{2i} \\
&&-ih_{\rho B_{c}D^{\ast}B^{\ast}}\left[ B_{c}^{-}(\partial _{\mu }D_{\nu }^{\ast })\vec{\tau} \cdot \vec{\rho}{%
_{\alpha }}B_{\beta }^{\ast }+B_{c}^{+}(\partial _{\mu }{\large \bar{B}}%
_{\nu }^{\ast }){\Large \rho }{_{\alpha }}\bar{D}_{\beta }^{\ast }%
\right]   \notag
\end{eqnarray}
\end{subequations}%
\noindent In Eqs. (\ref{eq2}) and (\ref{eq3}) $\overrightarrow{\tau}$ represents Pauli spin matrices, and $\overrightarrow{\pi}$ and $\overrightarrow{\rho}$ represent isospin triplets,
\begin{equation*}
\overrightarrow{\pi} = ( \pi _{1},\pi _{2},\pi _{3}),  \   \   \ \overrightarrow{\rho}= ( \rho _{1},\rho _{2},\rho _{3}),
\end{equation*}
\noindent while vector and pseudoscalar charm and bottom meson doublets are given as

\begin{equation*}
\begin{tabular}{lll}
$\bar{D}_{\mu }^{\ast }=(\bar{D}_{\mu }^{\ast 0}, D_{\mu }^{\ast -})^T$ ,  &  $\bar{D}= (\bar{D}^{0},  D^{-})^{T}$, & $D=(D^{0}, D^{+}),$  \\
$B_{\mu }^{\ast }=\left(B_{\mu }^{\ast +},  B_{\mu }^{\ast 0}  \right)^T$,  &$\bar{B}=\left(B^{-}, \bar{B}^{0 }\right)$,  & $B=(B^{+}, B^{0})^T$.
\end{tabular}
\end{equation*}

\subsection{\textbf{Amplitudes for $B_{c}$ meson dissociation}}
\noindent For calculating the cross section for $B_{c}$ meson dissociation by $\pi$ and $\rho$ mesons, we use the effective Lagrangian given in Eqs. (\ref{eq2})  and (\ref{eq3}). In this paper we are only reporting the scattering amplitudes of anomalous processes and of additional diagrams which are dependent on the anomalous couplings. Absorption amplitudes of other diagrams which depend only on normal couplings are given in Refs. \cite{lodhi2011, faisal2011}.
\noindent Diagrams of the process $B_{c}^++\pi \rightarrow D^{\ast}+B$ are shown in Fig. 1 (2a to 2c) and the amplitudes of the diagrams are
\begin{subequations}
\label{5}
\begin{eqnarray}
M_{2a} &=& g_{\pi D^{\ast }D^{\ast }}g_{B_{c}BD^{\ast }}\ \ {\large %
\varepsilon }_{\mu \nu \alpha \sigma }p_{3}^{\mu }(p_{3}-p_{1})_{\beta
}\frac{-i}{t-m_{D^{\ast }}^{2}}\left( g^{\alpha\sigma }-\frac{(p_{1}-p_{3})^{\alpha
}(p_{1}-p_{3})^{\sigma }}{m_{D^{\ast }}^{2}}\right)   \label{5a} \\
&&(-p_{2}-p_{4})^{\nu }{\large \varepsilon }^{{\beta }}_{D^{*}}(p_{3}), \notag \\
M_{2b} &=& g_{\pi BB^{\ast}}g_{B_{c}B^{\ast }D^{\ast }}\ {\large %
\varepsilon }_{\mu \nu \alpha \sigma } p_{3}^{\mu}
(p_{1}+p_{4})^{\nu}\frac{-i}{u-m_{B^{\ast }}^{2}}\left( g^{\alpha\sigma }-%
\frac{(p_{1}-p_{4})^{\alpha }(p_{1}-p_{4})^{\sigma }}{m_{B^{\ast }}^{2}}\right)
\label{5b} \\
&&(p_{3}-p_{2})_{\beta }{\large \varepsilon }^{{\beta }}_{D^{*}} (p_{3}), \notag \\
M_{2c} &=&-ig_{\pi B_{c}BD^{\ast }}\ {\large \varepsilon }_{\mu \nu \alpha
\beta } p_{1}^{\alpha }p_{4}^{\mu }p_{2}^{\nu }{\large \varepsilon }^{{\beta }}_{D^{*}} (p_{3}).  \label{5c}
\end{eqnarray}%
And the full amplitude is written as
\begin{equation}
M_{2}=M_{2a}+M_{2b}+M_{2c}.  \label{4d}
\end{equation}
\end{subequations}
\noindent Diagrams of the process $B_{c}^++\pi \rightarrow D+B^{\ast}$ are shown in Fig. 1 (3a to 3c) and the amplitudes of the diagrams are
\begin{subequations}
\label{5}
\begin{eqnarray}
M_{3a} &=&g_{\pi D^{\ast }D}g_{B_{c}B^{\ast }D^{\ast }}\ \ {\large %
\varepsilon }_{\mu \nu \alpha \sigma }p_{4}^{\mu}(p_{4}-p_{2})_{\beta }%
\frac{-i}{t-m_{D^{\ast }}^{2}}\left( g^{\alpha\sigma }-\frac{(p_{1}-p_{3})^{\alpha
}(p_{1}-p_{3})^{\sigma}}{m_{D^{\ast }}^{2}}\right)   \label{5a} \\
&&(p_{1}+p_{3})^{\nu}{\large \varepsilon }^{\beta}_{B^{*}}(p_{4}),  \notag \\
M_{3b} &=&g_{\pi B^{\ast }B^{\ast }}g_{B_{c}B^{\ast }D}\ {\large %
\varepsilon }_{\mu \nu \alpha \sigma }p_{4}^{\mu
}(p_{4}-p_{1})_{\beta}\frac{-i}{u-m_{B^{\ast }}^{2}}\left( g^{\alpha\sigma}-%
\frac{(p_{1}-p_{4})^{\alpha }(p_{1}-p_{4})^{\sigma}}{m_{D^{\ast }}^{2}}\right)
\label{5b} \\
&&(-p_{2}-p_{3})^{\nu}{\large \varepsilon }^{\beta}_{B^{*}}(p_{4}),  \notag \\
M_{3c} &=&-i g_{\pi B_{c}DB^{\ast }}\ \ {\large \varepsilon }_{_{\mu \nu
\alpha \beta }}p_{2}^{\alpha }p_{3}^{\mu }p_{1}^{\nu }{\large \varepsilon }^{\beta}_{B^{*}}(p_{4}). \label{5c}
\end{eqnarray}
And the full amplitude is written as
\begin{equation}
M_{3}=M_{3a}+M_{3b}+M_{3c}.
\label{5d}
\end{equation}
\end{subequations}

\noindent Diagrams of the process $ B_{c}^{+}+ \pi \rightarrow D^{\ast}+ B^{\ast}$ are shown in Fig. 1 (4a to 4e). The amplitudes of diagram 4d and 4e which depend on anomalous couplings are
\begin{subequations}
\label{6}
\begin{eqnarray}
M_{4d} &=& g_{\pi D^{\ast} D^{\ast}}g_{B_{c}B^{\ast }D^{\ast}} {\large \varepsilon }_{\sigma \lambda \alpha \beta}%
{\large \varepsilon }^{\sigma \lambda}_{\gamma \zeta } p_{4}^{\gamma
}(p_{3}-p_{1})_{\mu }\frac{-i}{t-m_{D^{\ast}}^2}   \label{6a} \\
&&\left( g^{\alpha\beta }-\frac{(p_{1}-p_{3})^{\alpha }(p_{1}-p_{3})^{\beta}}{%
m_{D^{\ast}}^{2}}\right) p_{3}^{\zeta }(p_{4}-p_{2})_{\nu }\varepsilon
_{D^{\ast }}^{\mu }(p_{3})\varepsilon _{B^{\ast}}^{\nu }(p_{4}),  \notag \\
M_{4e} &=&g_{\pi B^{\ast}B^{\ast}}g_{B_{c}D^{\ast}B^{\ast }}{\large \varepsilon }%
_{\sigma \lambda \alpha \beta }{\large \varepsilon }^{\sigma \lambda}
_{\gamma \zeta }(p_{4}-p_{1})_{\nu }p_{3}^{\gamma } \frac{-i}{u-m_{B^{\ast}}^{2}}  \label{6b} \\
&&\left( g^{\alpha\beta }-\frac{(p_{1}-p_{4})^{\alpha }(p_{1}-p_{4})^{\beta }}
{m_{B^{\ast}}^{2}}\right) (p_{3}-p_{2})_{\mu }p_{4}^{\zeta }\varepsilon
_{D^{\ast}}^{\mu }(p_{3})\varepsilon _{B^{\ast}}^{\nu }(p_{4}).  \notag
\end{eqnarray}
\begin{figure}[H]
\begin{center}
\includegraphics[angle=0,width=0.38\textwidth]{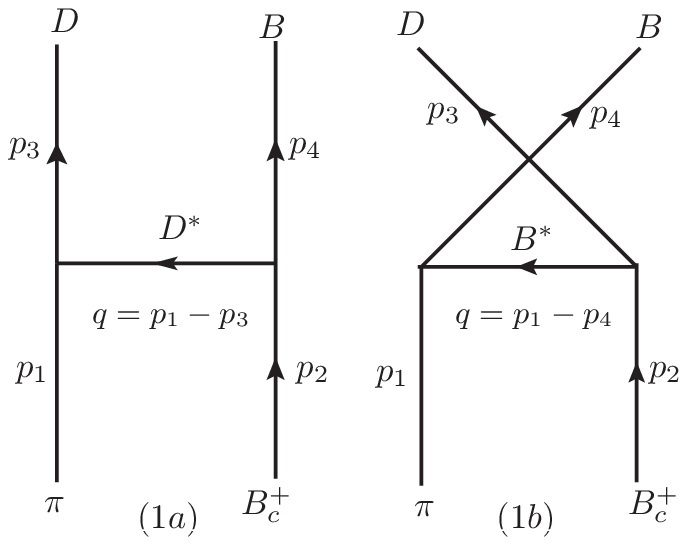}\bigskip\\
\includegraphics[angle=0,width=0.55\textwidth]{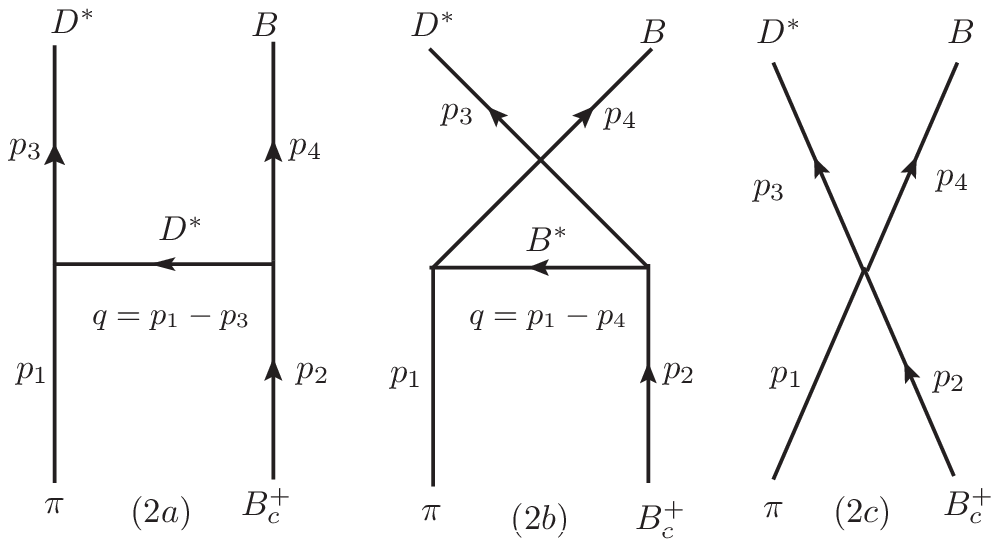}\bigskip\\
\includegraphics[angle=0,width=0.55\textwidth]{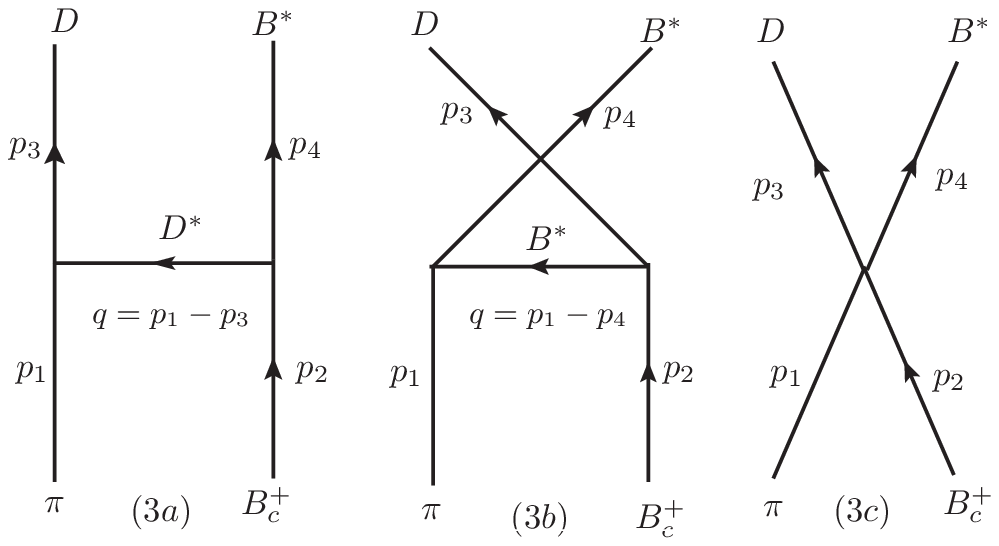}\bigskip\\
\includegraphics[angle=0,width=0.9\textwidth]{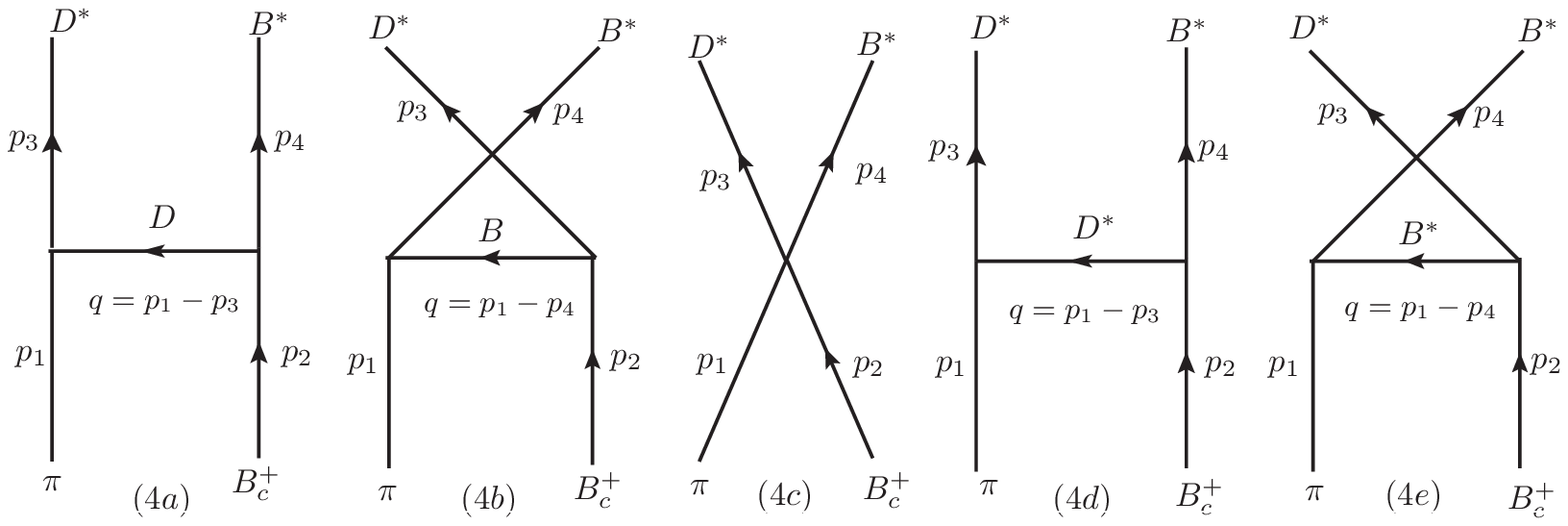}
\end{center}
\caption{Feynman diagrams of $B_{c}$ dissociation processes $(1)\; B_{c}^{+}+\pi\rightarrow D+B$, $ (2)\; B^{+}_{c}+\pi\rightarrow D^{\ast }+B$ ,
$ (3)\; B^{+}_{c}+\pi\rightarrow D+B^{\ast }$ and $ (4)\; B^{+}_{c}+\pi\rightarrow D^{\ast }+B^{\ast }$, respectively.} \label{Fig1}
\end{figure}%
\noindent And the full amplitude is written as
\begin{equation}
M_{4}=M_{4a}+M_{4b}+M_{4c}+M_{4d}+M_{4e}.  \label{6c}
\end{equation}%
\end{subequations}
Now we report the absorption amplitudes of the anomalous processes of $B_{c}$ by $\rho$. Diagrams of the process $B_{c}^{+}+\rho \rightarrow D+B$ are shown in
Fig. 2 (5a to 5c). The amplitudes of these diagrams are
\begin{subequations}
\label{7}
\begin{eqnarray}
M_{5a} &=&g_{B_{c}BD^{\ast }}g_{\rho D^{\ast }D}{\large \varepsilon }_{\sigma
\nu \alpha \beta}p_{1}^{\nu}   \left(-p_{2}-p_{4}\right)^{\sigma}\frac{-i}{t-m_{D^{\ast }}^{2}}  \label{7a} \\
&&\left( g^{\alpha \beta }-\frac{\left( p_{1}-p_{3}\right) ^{\alpha }\left(
p_{1}-p_{3}\right)^{\beta}}{m_{D^{\ast }}^{2}}\right) (p_{3}-p_{1})_{\mu}\varepsilon _{\rho}^{\mu }(p_{1}),  \notag \\
M_{5b} &=&g_{\rho B^{\ast }B}g_{B_{c}B^{\ast }D}{\large \varepsilon }%
_{\sigma \nu \alpha \beta}p_{1}^{\sigma }\frac{-i}{u-m_{B^{\ast }}^{2}} \left(
-p_{3}-p_{2}\right)^{\nu}\label{7b} \\
&&\left( g^{\alpha\beta }-\frac{\left( p_{1}-p_{4}\right) ^{\alpha }\left(
p_{1}-p_{4}\right) ^{\beta }}{m_{B^{\ast }}^{2}}\right)\left(
p_{4}-p_{1}\right)_{\mu} \varepsilon _{\rho}^{\mu }(p_{1}),  \notag \\
M_{5c} &=&-ig_{\rho B_{c}BD{\large}}{\large \varepsilon }_{\mu \nu
\alpha \beta}p_{2}^{\nu }p_{3}^{\alpha }p_{4}^{\beta }\varepsilon
_{\rho}^{\mu }(p_{1}).  \label{7c}
\end{eqnarray}%

And the full amplitude is written as
\begin{equation}
M_{5}=M_{5a}+M_{5b}+M_{5c}.  \label{7d}
\end{equation}%
\end{subequations}

\noindent Diagrams of the process $B_{c}^{+}+\rho \rightarrow D^{\ast}+B$ are shown in Fig. 2 (6a to 6d). The amplitudes of the anomalous diagram 6d is given as
\begin{subequations}
\label{8}
\begin{eqnarray}
M_{6d} &=&g_{\rho B^{\ast }B}g_{B_{c}B^{\ast }D^{\ast }}{\large \varepsilon }%
_{\delta \gamma \sigma \lambda}{\large \varepsilon }^{\delta \gamma}
_{\alpha \beta}p_{1}^{\sigma }p_{4}^{\lambda }\left( p_{3}-p_{1}\right) _{\mu }%
\frac{-i}{t-m_{B^{\ast }}^{2}}  \label{8a} \\
&&\left( g^{\alpha \beta }-\frac{\left( p_{1}-p_{3}\right) ^{\alpha }\left(
p_{1}-p_{3}\right) ^{\beta }}{m_{B^{\ast }}^{2}}\right) \left(
p_{4}-p_{2}\right)_{\nu }  \varepsilon _{\rho}^{\mu }(p_{1}) \varepsilon _{D^{\ast}}^{\nu }(p_{4}). \notag
\end{eqnarray}
And the full amplitude is written as
\begin{equation}
M_{6}=M_{6a}+M_{6b}+M_{6c}+M_{6d}.  \label{8b}
\end{equation}
\end{subequations}
\noindent Diagrams of the process $B_{c}^{+}+\rho \rightarrow D+B^{\ast}$ are shown in Fig. 2 (7a to 7d). The amplitude of the anomalous diagram 7d is given as
\begin{subequations}
\label{9}
\begin{eqnarray}
M_{7d} &=&g_{\rho D^{\ast}D}g_{B_{c}B^{\ast }D^{\ast }}{\large \varepsilon}%
_{\sigma \lambda \gamma \delta}{\large \varepsilon }^{\gamma \delta} _{\alpha \beta}
p_{1}^{\sigma }p_{4}^{\lambda }\left( p_{3}-p_{1}\right)_{\mu}%
\;\frac{-i}{t-m_{D^{\ast }}^{2}}  \label{9a} \\
&&\left( g^{\alpha \beta }-\frac{\left( p_{1}-p_{3}\right) ^{\alpha }\left(
p_{1}-p_{3}\right) ^{\beta }}{m_{D^{\ast }}^{2}}\right) \left(
p_{2}-p_{4}\right)_{\nu }\;\varepsilon _{\rho}^{\mu }(p_{1})\varepsilon
_{D^{\ast}}^{\nu }(p_{4}).  \notag
\end{eqnarray}%
And the full amplitude is written as
\begin{equation}
M_{7}=M_{7a}+M_{7b}+M_{7c}+M_{7d}.  \label{9b}
\end{equation}%
\end{subequations}
\begin{figure}[H]
\begin{center}
\includegraphics[angle=0,width=0.55\textwidth]{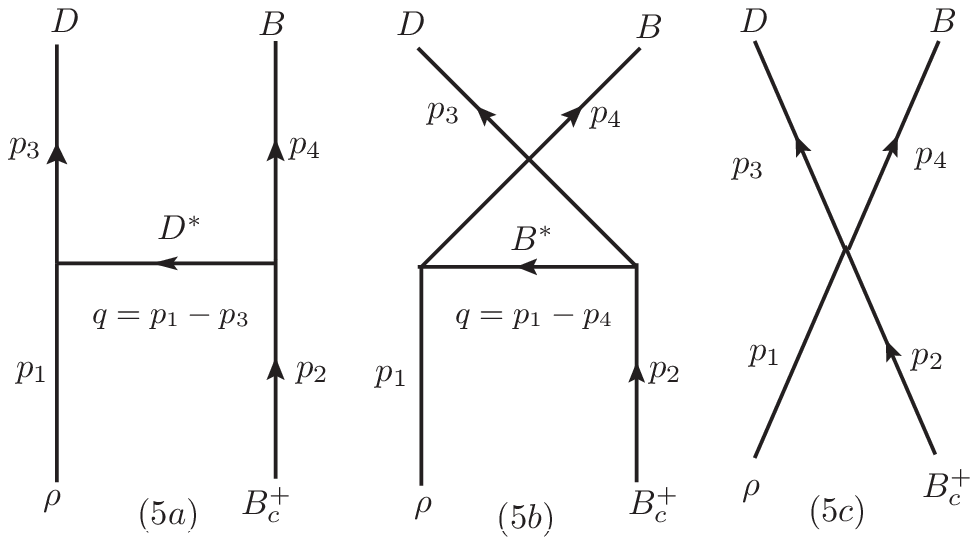}\bigskip\\
\includegraphics[angle=0,width=0.75\textwidth]{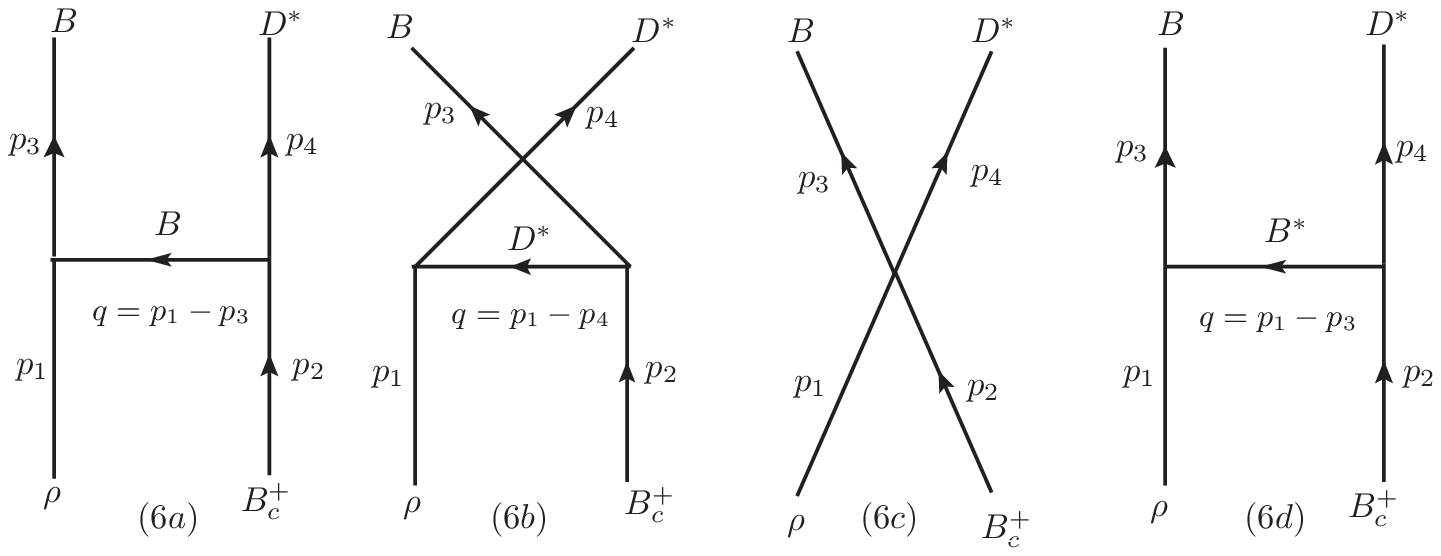}\bigskip\\
\includegraphics[angle=0,width=0.75\textwidth]{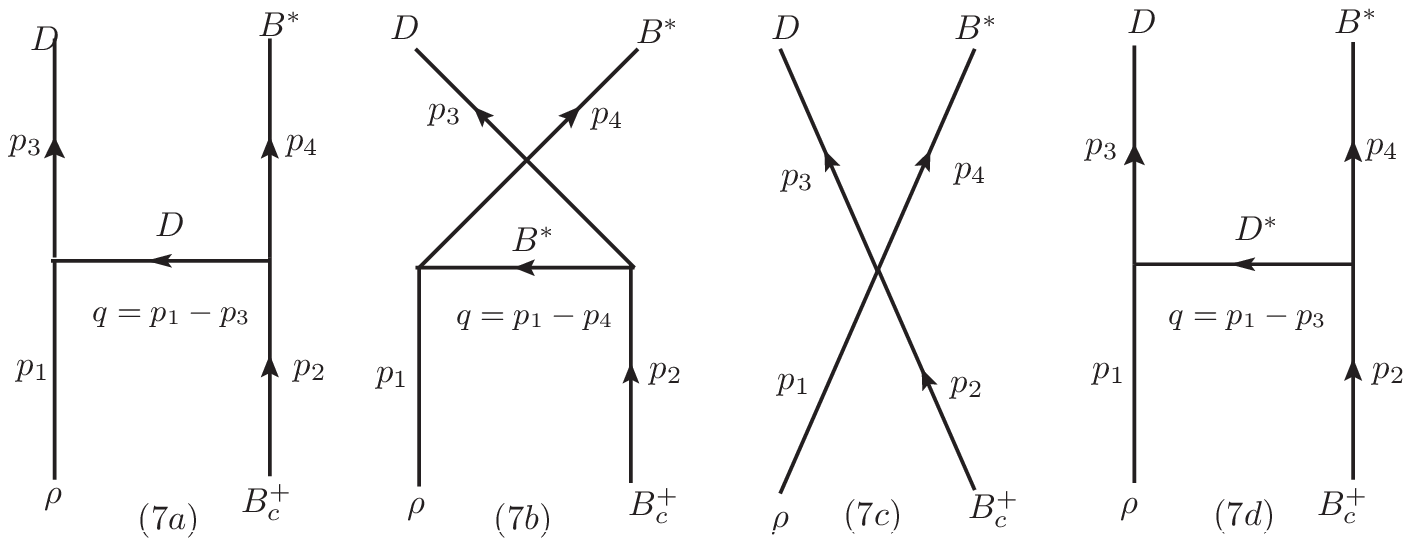}\bigskip\\
\includegraphics[angle=0,width=0.95\textwidth]{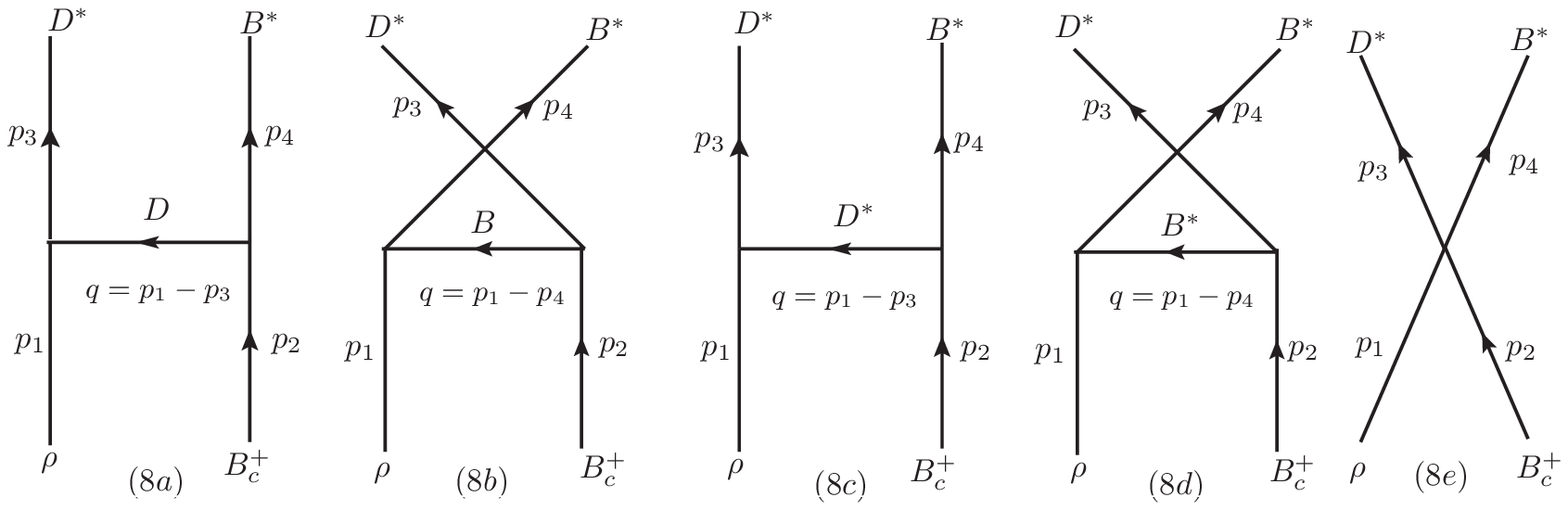}
\end{center}
\caption{Feynman diagrams of $B_{c}$ dissociation processes $ (5)\; B^{+}_{c}+\rho\rightarrow D+B$, $ (6) \;B_{c}^{+}+\rho\rightarrow D^{\ast}+B$,
$ (7) \; B^{+}_{c}+\rho\rightarrow D+B^{\ast }$ and $ (8)\; B^{+}_{c}+\rho\rightarrow D^{\ast}+B^{\ast}$.} \label{Fig2}
\end{figure}
\newpage
\noindent Diagrams of the process $B_{c}^{+}\rho \rightarrow D^{\ast}+B^{\ast}$ are shown in Fig. 2 (8a to 8e), the amplitudes of these diagrams are
\begin{subequations}
\label{10}
\begin{eqnarray}
M_{8a} &=&g_{_{\rho D^{\ast} D}}g_{_{B_{c}B^{\ast }D}}
{\textbf{\large}\varepsilon }_{\mu \nu \alpha \beta}p_{1}^{\alpha }
\left(p_{3}-p_{1}\right) ^{\beta }\frac{i}{t-m_{D}^{2}}\left(
p_{4}-2p_{2}\right)_{\lambda}\varepsilon _{\rho}^{\mu}(p_{1})  \label{10a}
\\ &&\varepsilon _{D^{\ast}}^{\nu }(p_{3})\varepsilon _{B^{\ast}}^{\lambda }(p_{4}),\notag \\
M_{8b} &=& - g_{_{\rho B B^{\ast }}}g_{_{B_{c}B D^{\ast}}}
{\large \varepsilon }_{\mu \lambda \alpha \beta}p_{1}^{\alpha }p_{4}^{\beta }
\frac{i}{u-m_{B}^{2}}\left( p_{3}-2p_{2}\right)_{\nu }\varepsilon
_{\rho}^{\mu }(p_{1})\varepsilon _{D^{\ast}}^{\nu }(p_{3})
\varepsilon _{B^{\ast}}^{\lambda}(p_{4}),  \label{10b} \\
M_{8c} &=&g_{_{\rho D^{\ast }D^{\ast}}}g_{_{B_{c}B^{\ast }D^{\ast}}}%
{\large \varepsilon }^{\sigma} _{\delta \alpha \beta}p_{4}^{\delta }\frac{-i}{%
t-m_{D^{\ast }}^{2}}[\left( 2p_{3}-p_{1}\right) _{\mu }g_{\sigma \nu }+\left(
2p_{1}-p_{3}\right) _{\nu }g_{\mu \sigma }  \label{10c} \\
&&+\left( -p_{3}-p_{1}\right) _{\sigma }g_{\mu \nu }]\left( g^{\alpha \beta }-%
\frac{\left( p_{1}-p_{3}\right) ^{\alpha }\left( p_{1}-p_{3}\right) ^{\beta }%
}{m_{D^{\ast }}^{2}}\right) \left( p_{4}-p_{2}\right)_{\lambda }  \notag \\
&&\varepsilon _{\rho}^{\mu }(p_{1})\varepsilon _{D^{\ast}}^{\nu }(p_{3})\varepsilon
_{{B^{\ast}}}^{\lambda }(p_{4}),  \notag \\
M_{8d} &=&g_{_{\rho B^{\ast}B^{\ast}}}g_{_{B_{c}B^{\ast }D^{\ast }}}{\large %
\varepsilon }_{\delta \alpha \beta }^{\sigma} p_{3}^{\delta} \frac{-i}{u-m_{B^{\ast }}^{2}}[\left(
-2p_{4}+p_{1}\right) _{\lambda }g_{\sigma\mu }+\left( p_{1}+p_{4}\right)
_{\sigma }g_{\mu \lambda }  \label{10d} \\
&&+\left( p_{4}-2p_{1}\right) _{\mu }g_{\sigma \lambda }]\left( g^{\alpha
\beta }-\frac{\left( p_{1}-p_{4}\right) ^{\alpha }\left( p_{1}-p_{4}\right)
^{\beta }}{m_{B^{\ast }}^{2}}\right) \left(p_{2}- p_{3} \right)_{\nu } \notag \\
&&\varepsilon _{\rho}^{\mu }(p_{1})\varepsilon _{D^{\ast}}^{\nu }(p_{3})\varepsilon
_{B^{\ast}}^{\lambda }(p_{4}),  \notag \\
M_{8e} &=&(-i g_{\rho B_{c}B^{\ast}D^{\ast}}{\large \varepsilon } _{\mu \nu \lambda\beta}p^{\beta}_{2}
+ih_{_{\rho B_{c}B^{\ast }D^{\ast }}}{\large \varepsilon }_{\mu \nu \lambda \beta}p^{\beta}_{4})
\varepsilon _{\rho}^{\mu}(p_{1})\varepsilon _{D^{\ast}}^{\nu }(p_{3})\varepsilon _{B^{\ast}}^{\lambda }(p_{4}).
\label{10e}
\end{eqnarray}%
And the full amplitude is written as
\begin{equation}
M_{8}=M_{8a}+M_{8b}+M_{8c}+M_{8d}+M_{8e}.  \label{10f}
\end{equation}%
\end{subequations}\\
We define the four-momenta of the incoming particles as $p_{1}$ and $p_{2}$ and those of the final particles as $p_{3}$ and $p_{4}$, which then defines $t=(p_{1}-p_{3})^{2}$ and $s=(p_{1}+p_{2})^{2}$. Here $m_{D}$, $m_{D^{\ast }}$, $m_{B}$ and  $m_{B^{\ast}}$ represent the $D$, $D^{\ast }$, $B$ and $B^{\ast }$ mesons masses, respectively. The polarization vector of a vector meson with momentum $p_{i}$ is represented by $\large\varepsilon_{i}(p_{i})$. After averaging (summing) over initial (final) spins and including isospin factor, we calculate the cross sections by using the total amplitudes specified in above equations. The isospin factor for calculating these cross section is 2 for all the processes.

\section{Dissociation Cross-Sections of $B_{c}$ Meson}
\subsection{Numerical values of input parameters}

\begin{table}
\begin{center}
\begin{tabular}{|c|c|c|}
\hline
Coupling constant & Value & Method of Derivation \\\hline
$g_{\pi D^*D^*}$     & 9.08 $\text{GeV}^{-1}$ & Heavy quark symmetries  \\
$g_{\pi B^*B^*}$     & 2.34  $\text{GeV}^{-1}$ & Heavy quark symmetries  \\
$g_{B_cB^*D^*}$      & 6.134 $\text{GeV}^{-1}$ & Heavy quark symmetries \\
$g_{\pi B_cD^*B}$    & 21.56 $\text{GeV}^{-3}$ & $SU(5)$ symmetry \\
$g_{\pi B_cDB^*}$    & 21.56 $\text{GeV}^{-3}$ & $SU(5)$ symmetry \\\hline
$g_{\rho D^*D}$      & 2.82  $\text{GeV}^{-1}$ & VMD\\
$g_{\rho B^*B}$      & 2.58  $\text{GeV}^{-1}$ & Heavy quark symmetries \\
$g_{\rho B_cBD}$     & 21.56 $\text{GeV}^{-3}$ & $SU(5)$ symmetry \\\
$g_{\rho B_cD^*B^*}$ & 67    $\text{GeV}^{-1}$ & $SU(5)$ symmetry\\\hline
\end{tabular}
\end{center}
\caption{Coupling constants for anomalous interactions $B_{c}$ with $\pi$ and $\rho$ mesons.}
\label{table1}
\end{table}
\noindent Numerical values of all the meson masses are taken from Particle Data Group \cite{pdg}. Estimation of the coupling constants of effective Lagrangian \label{2} is required for calculating the cross sections. To fix the couplings for the normal processes, we follow the methods of Refs. \cite{lin2000,haglin2001}; we refer to Ref. \cite{lin2000} for details. In a similar way we have determined the couplings for the anomalous interactions, which are reported in this paper whereas normal couplings are given in Refs. \cite{lodhi2011,faisal2011}. The coupling ${\large g}_{D^{\ast }D^{\ast }\pi }$ which has a dimension of $\textmd{GeV}^{-1}$ is fixed by applying the heavy quark spin symmetry. We follow Ref. \cite{oh2001} in which this coupling is given as\\
\begin{equation}
{\large g}_{D^{\ast}D^{\ast}\pi}=\frac{\large g_{D^{\ast}D\pi}}{\overline{M}_{D}}\approx 9.08\textmd{ GeV}^{-1}
\end{equation}%
\noindent where $\overline{M}_{D}$ represents the average mass of $D$ and $ D^{\ast }$. \\
\noindent For ${\large g}_{\rho D^{\ast}D}$ couplings, we can apply the VMD (Vector Meson Dominance) model \cite{lin2000} to the radiative decays of $D^{\ast}$ into $D$, i.e., $D^{\ast } \rightarrow D \gamma$. We use the same method as in ref. \cite{oh2001}; this leads to
\begin{equation}
{\large g}_{\rho D^{\ast }D}{\large =2.82\textmd{ GeV}}^{-1}.
\end{equation} \\
\noindent The coupling constants ${\large g}_{\rho B^{\ast} B}$,
${\large g}_{B_{c}B^{\ast}D^{\ast}}$,
$ {\large g}_{\pi B^{\ast}B^{\ast}}$ can be approximated by
$\frac{{\large\ g}_{\rho B B}}{{\large \overline{M}}_{B}}$,
$\frac{{\large g}_{B_{c}B^{\ast }D}}{{\large \overline{M}}_{D}}$ and
$\frac{{\large g}_{\pi B^{\ast }B}}{{\large \overline{M}}_{B}}$,
respectively, in heavy quark mass limit as in Ref. \cite{chan1997}.
\noindent Since no experimental or phenomenological information is available on the 4-point vertices. In this case we use SU(5) symmetry relations to relate a 4-point coupling to the product of two 3-point couplings and assumes that the symmetry breaking effects in the 4-point coupling constants are included via phenomenological values of the 3-point couplings, as argued in Ref. \cite{lin2000}. Hence, using the symmetry relations and the phenomenological estimates of the 3-point vertices, as given in Refs. \cite{lodhi2011,faisal2011} and as given above, we have
\begin{equation}
h_{\rho B_{c}D^{\ast}B^{\ast}}=g_{\rho B_{c}D^{\ast}B^{\ast}}= 2 g_{\rho D^{\ast}D} g_{B_{c}B D^{\ast}}\approx 67{\large \textmd{GeV}}^{-1}.\\
\label{hcoup}
\end{equation}
\noindent However, for $g_{\pi B_{c} B^{\ast}D}$, $g_{\pi B_{c} D^{\ast}B}$ and $g_{\rho B_{c}D B}$ couplings, it is not possible to write these couplings as a product of two 3-point coupling constants because of the difference in their dimensions. Hence, in this can we directly use SU(5) symmetry relation assuming  the symmetry breaking effects change $F_{\pi}$ to $F_{D}$ \cite{Pari1991}, where $F_{\pi}$ and $F_{D}$ are pion and $D$ mesons decay constants respectively. Here we have used $F_{D}\approx 2.3F_{\pi}$ as in Ref. \cite{oh2001}. This gives
\begin{equation}
g_{\pi B_{c}DB^{\ast}}=g_{\pi B_{c}D^{\ast}B}=g_{\rho B_{c}D B} =\frac{g_{B_{c}DB^{\ast}} N_{c}}  {6\pi ^{2}F_{D}^{3}}\approx 21.56{\large \textmd{GeV}}^{-3}.\\
\end{equation} \\
\noindent The three point coupling $g_{B_{c}D B ^{\ast}}$ is given in Refs. \cite{lodhi2011,faisal2011} and the constant factor of the couplings in the effective Lagrangian is given in Ref. \cite{oh2001}. We summarize the values of the coupling constants and methods for obtaining them in Table \ref{table1}.\\

\subsection{$B_{c}$ dissociation cross sections}

In the effective Hadronic Lagrangian the Hadrons represent the fundamental degrees of freedom. This treatment needs to be corrected by inclusion of form factors as the Hadrons are not the fundament particles and have finite sizes. The resulting changes in the transition amplitudes of any diagram can be accounted for by multiplying with the form factors of the interaction vertices involved in it. In this paper we have used the same monopole form factor as given in Refs. \cite{lodhi2011,faisal2011,faisal2012} to multiply with three point vertices of all the processes.\\
\begin{equation}
f_{3}=\frac{\Lambda ^{2}}{\Lambda ^{2}+\overline{\textbf{q}}^{2}}.
\end{equation}
\noindent Here, $\Lambda $ represents a cutoff parameter and square of the exchange three momentum for the system in c.m (centre of mass) frame is represented by $\overline{\textbf{q}}^{2}$. Where $\overline{\textbf{q}}^{2}= (\textbf{p}_{1}- \textbf{p}_{3})^{2}$ for t channel diagrams and $(\textbf{p}_{1}-\textbf{p}_{4})^{2}$ for u channel diagrams. This form was used to calculate the cross sections of $B^{+}_{c}$ by $\pi$, $\rho$ mesons and nucleons in Refs. \cite{lodhi2007,lodhi2011,faisal2011,faisal2012} and also in Refs. \cite{lin2000,lin2001,liu2006} to calculate the hadronic cross sections of charmonium, bottomonium, and eta mesons.\\
\noindent Following form factor is used at four point vertices of all the processes.\\
\begin{equation}
f_{4}=\left( \frac{\Lambda ^{2}}{\Lambda ^{2}+\overline{\textbf{q}}^{2}}\right) ^{2},
\end{equation}
\noindent where $\overline{\textbf{q}}^{2}=\frac{1}{2}\left[ (\textbf{p}_{1}- \textbf{p}_{3})^{2}+(\textbf{p}_{1}-\textbf{p}_{4})^{2}\right] _{c.m}$. Generally, cutoff parameter may take different values for different vertices. In some cases cutoff parameters of the form factors used with meson or baryon exchange models can be fitted to experimental data of hadronic cross sections \cite{machleid1987}. In the absence of any experimental data, we may provide an estimate on the basis of size of the interacting hadrons. It is shown in Ref. \cite{lodhi2011} that a variation in the range 1.2 to 1.8 GeV is consistent with the known sizes of the interacting hadrons. As in the previous studies \cite{lodhi2011,faisal2011,oh2001} and also based on the results given in Ref. \cite{yasui2009}, we consider the same cutoff parameters for all the processes and use two values $1$ and $2$ GeV.

\noindent Figs. 3(a-d) show the cross sections for $B_{c}$ dissociation with and without form factor $(f_{3})$ for the processes (a) $B_{c}^{+}+\pi\rightarrow D+B$,
(b) $B^{+}_{c}+\pi \rightarrow D^{\ast }+B$ , (c) $ B^{+}_{c}+\pi\rightarrow D+B^{\ast }$, and (d) $B^{+}_{c}+\pi\rightarrow D^{\ast }+B^{\ast }$ as a function of total c.m energy $\sqrt{s}$. Cross sections with and without form factors are represented by solid and dashed curves respectively. Lower and upper dashed curves are with cutoff parameters $\Lambda = 1$ and 2 GeV respectively. It can be seen that including the form factors substantially suppress the cross sections. The cross sections remain increasing rapidly at threshold for all four processes. The process (a) $B^{+}_{c}+\pi \rightarrow D+B$ is a normal process and it does not include any anomalous diagram. Threshold energy of this process is 7.15 GeV. It can be seen from Fig. 3a that for this process the cross section decreases as c.m energy increases and beyond 12 GeV it becomes very small with the form factors included. The same plot is also reported in our previous work without isospin average factor \cite{lodhi2011}. Both (b) $B^{+}_{c}+\pi \rightarrow D^{\ast }+B$ and (c) $\ B^{+}_{c}+\pi \rightarrow D+B^{\ast}$ are anomalous processes.
 \begin{figure}[H]
\begin{center}
\includegraphics[angle=0,width=1.0\textwidth]{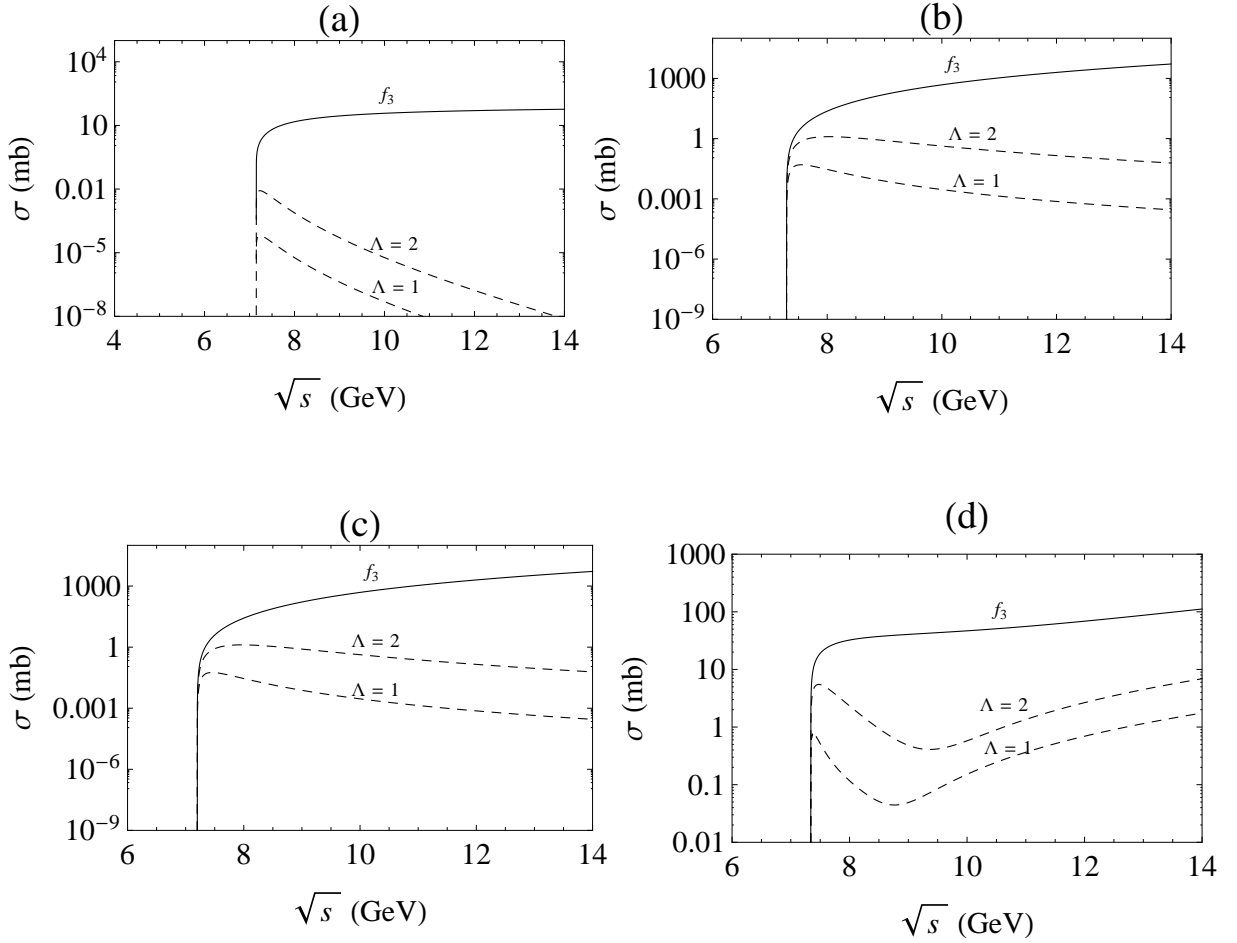}
\end{center}%
\caption{Cross section for the $B_{c}$ dissociation processes (a) $B_{c}^{+}\pi\rightarrow DB$,  (b) $B^{+}_{c}\pi\rightarrow D^{\ast }B$ , (c) $B^{+}_{c}\pi\rightarrow DB^{\ast }$, and (d) $B^{+}_{c}\pi\rightarrow D^{\ast }B^{\ast }$ respectively.}
\label{Fig.3}
\end{figure}

\noindent It can be seen from Fig. 3b that for the 2nd process the cross section ranges between 0.005 to 0.1 mb away from the threshold with the form factor included and its threshold energy is 7.29 GeV. Fig. 3c shows that for the 3rd process the cross section roughly varies between 0.0007 to 0.1 mb and threshold energy is 7.20 GeV. In the 4th process additional anomalous diagrams are included which are shown in Fig. 1 as diagrams 4d and 4e. This cross section was reported in our previous work without including the anomalous diagrams \cite{lodhi2011}. Threshold energy of this process is 7.29 GeV. Fig. 3d shows that the cross section ranges between 2 to 8 mb away from the threshold including the form factors.\\

 \begin{figure}[H]
\begin{center}
\includegraphics[angle=0,width=1.0\textwidth]{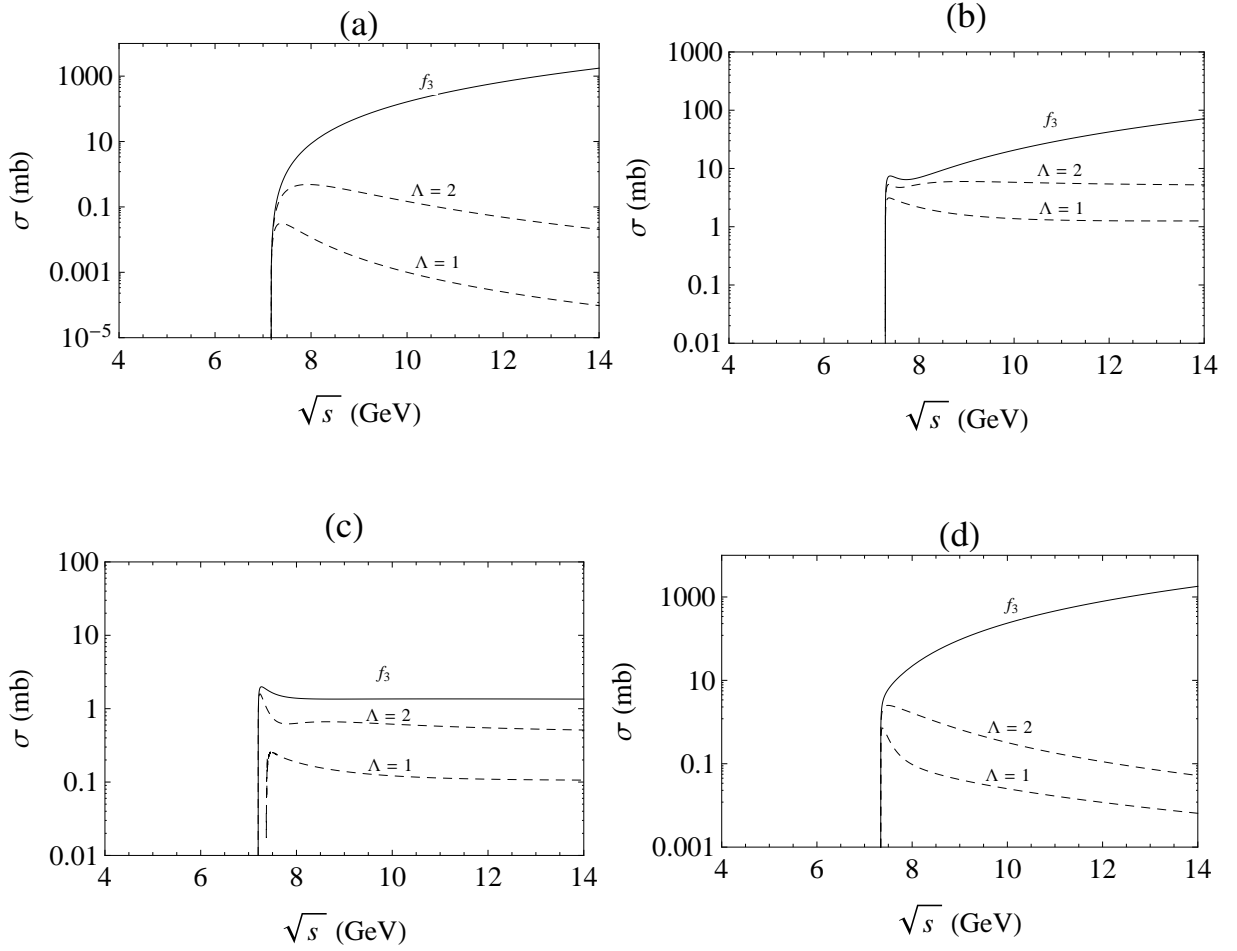}
\end{center}%
\caption{Cross section for the $B_{c}$ dissociation processes
(a) $B^{+}_{c}\rho \rightarrow DB$,
(b) $B_{c}^{+}\rho\rightarrow D^{\ast}B$,
(c) $B^{+}_{c}\rho\rightarrow DB^{\ast }$, and
(d) $B^{+}_{c}\rho\rightarrow D^{\ast}B^{\ast}$ respectively.}
\label{Fig.4}
\end{figure}
\noindent Figs. 4(a-d) shows the cross sections for $B_{c}$ dissociation with and without form factor $(f_{3})$ for the processes (a) $B_{c}^{+}+\rho\rightarrow D+B$,
(b) $B_{c}^{+}+\rho\rightarrow D^{\ast }+B$, (c) $B_{c}^{+}+ \rho\rightarrow D+B^{\ast }$, and (d) $B_{c}^{+} +\rho\rightarrow D^{\ast }+B^{\ast }$ respectively as a function of the total c.m energy $\sqrt{s}$. The cross sections again increase rapidly at the threshold  for all four processes. As shown in Fig. 4a that for the first process the cross section ranges from 0.0001 to 0.03 mb when form factor is included. Fig. 4d shows that for the 4th process the cross section with form factor approximately ranges between 0.01 to 0.08 mb for large $\sqrt{s}$. It is noted that both 1st and 4th are anomalous processes where as for 2nd and 3rd processes additional diagrams are introduced by anomalous interactions. Previously in Ref. \cite{faisal2011} we have studied these two processes without including the anomalous diagrams. As shown in Fig. 4b for the 2nd process the cross section with form factor ranges between 2 to 7 mb and threshold energy is 7.20 GeV. Fig. 4c shows that for the 3rd process the cross section roughly ranges between 0.1 to 0.7 mb and threshold energy for this process is 7.34 GeV.%
\subsection{Thermal average $B_{c}$ meson cross-sections}
The following formula can be used to calculate the thermal average cross-section \cite{liu2006}
\begin{equation}
\label{12}
\left\langle \sigma \upsilon \right\rangle =\left[ 4\alpha _{1}^{2}K_{2}\left(
\alpha _{1}\right) \alpha _{2}^{2}K_{2}\left( \alpha _{2}\right) \right]
^{-1}\times \underset{z_{0}}{\overset{\infty }{\int }} dz\left[ z^{2}-\left( \alpha _{1}+\alpha _{2}\right) ^{2}%
\right] \left[ z^{2}-\left( \alpha _{1}-\alpha _{2}\right) ^{2}\right]\\
K_{1}\left( z\right) \sigma \left( s=z^{2}T^{2}\right)
\end{equation}%
\strut with $\alpha _{i}=m_{i}/T, z_{0}=\textmd{max}(\alpha _{1}+\alpha _{2},\alpha_{3}+\alpha _{4})$, $K_{1}$ and $K_{2}$ are the modified Bessel functions
of second kind of order 1 and 2 respectively, $\upsilon$ is the relative velocity of initial particles and $T$ is the temperature of the hadronic matter. We have calculated the thermal average cross-sections of $B_{c}$ mesons including the anomalous parity interactions with form factor, as a function of temperature $T$. Fig. 5 shows the temperature dependence of thermal average cross sections with form factor for all 4 processes of $B_{c}$ interaction by $\pi$. The range of the temperature is taken from $0.1$ to $0.3$ GeV. Lower and upper dashed curves represent the values for $\Lambda =1$ and $2$ GeV respectively. The figure shows that contribution of first process is significantly small as compared to other three processes. In Fig. 6 thermal average cross sections for $B_{c}$ meson  absorption by $\rho$ mesons with form factor are given. Thermal average cross sections reported here are used to study yield of $B_{c}$ mesons in hadronic matter in the next section.
\section{Absorption rate of $B_{c}$ in RHIC}

Now we examine the effect of interactions of $B_{c}$ meson with the comovers on its absorption rate in the hot hadronic matter. The time evolution of its abundance is studied at RHIC energies using a schematic expanding fireball model with an initial $B_c$ abundance determined by the statistical model.

\subsection{Time evolution of $B_{c}$ Mesons}

Time evolution of $B_{c}$ meson density in hot hadronic matter can be studied by the rate equation expressed as
\begin{equation}
\partial _{\mu }(n_{B_{c}}u^{\mu})=\Psi,
\label{rate eq0}
\end{equation}
where ${\Psi }$ is composed of the source (the processes in which $B_{c}$ mesons are created) and/or the sink (the processes in which $B_{c}$ mesons are absorbed) terms, $u^{\mu }=\gamma (1,\mathbf{v})$ is the four velocity and is specified in term of fluid velocity $(\textbf{v})$ of the hadronic matter and Lorentz factor $\gamma $ \cite{liu2006}, and $n_{B_{c}}$ is the density of $B_{c}$ mesons.
\\
\noindent ${\Psi }$ as a source term is represented by $\Psi_{1}$ which is given as

\begin{equation}
 \Psi_{1} = \underset{a,b,c}{\sum }\langle\sigma_{bc\rightarrow B_{c}a}v_{bc}\rangle n_{b}n_{c}.
\end{equation}

\noindent ${\Psi }$ as a sink term is represented by $\Psi_{2}$ which can be written as
\begin{equation}
\Psi _{2}=\underset{a,b,c}{\sum }\langle\sigma _{B_{c}a\rightarrow bc}v_{B_{c}a}\rangle n_{B_{c}}n_{a},
\end{equation}
\noindent where $n_{a},n_{B_{c}},n_{b}$ and $n_{c}$ represent the densities of $a,B_{c}, b$ and $c$ mesons.

\begin{figure}[H]
\begin{center}
\includegraphics[angle=0,width=0.90\textwidth]{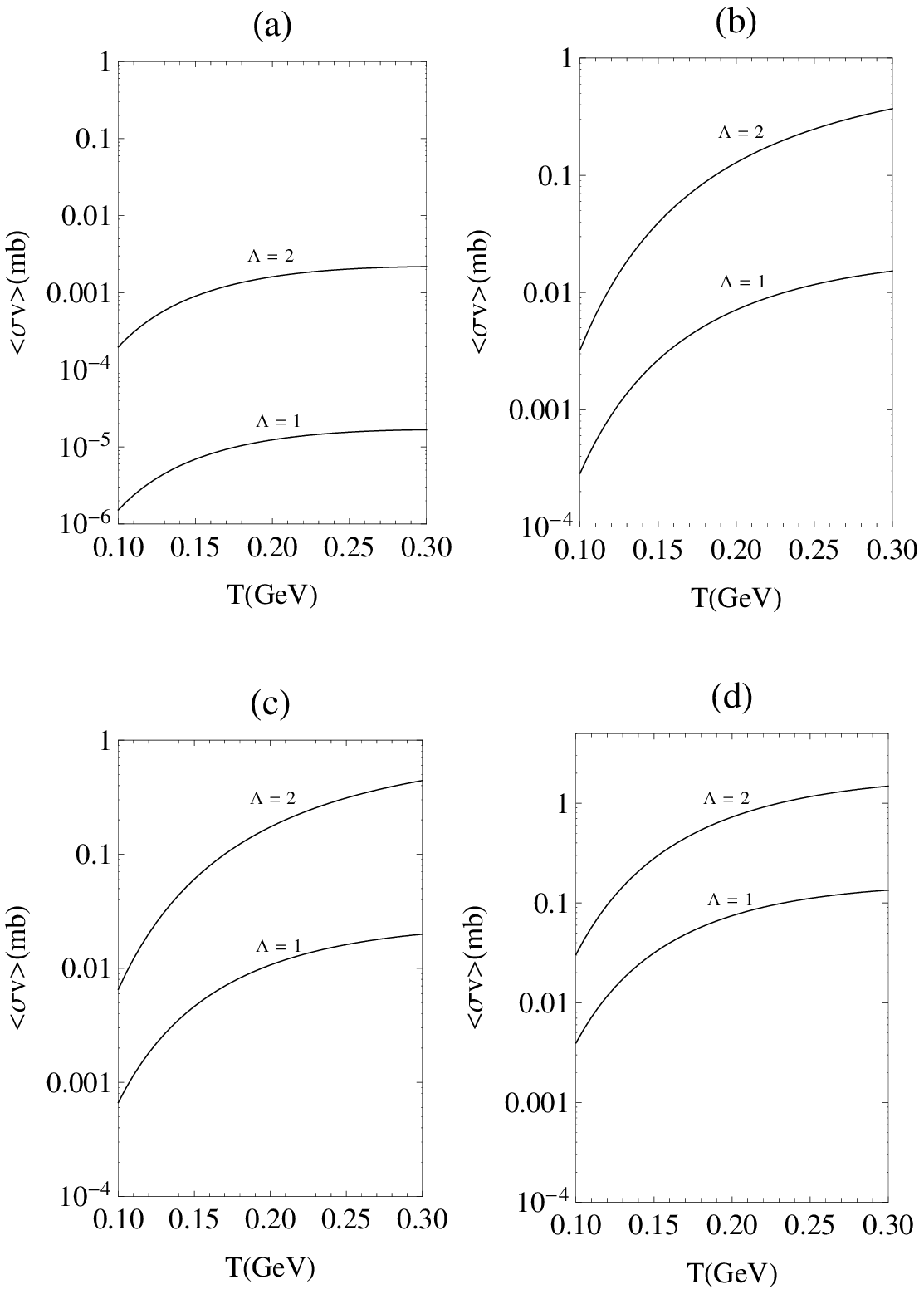}
\end{center}%
\caption {Thermal average $B_{c}$ dissociation process
(a) $B_{c}^{+}\pi\rightarrow DB$,
(b) $B^{+}_{c}\pi\rightarrow D^{\ast }B$ ,
(c) $B^{+}_{c}\pi\rightarrow DB^{\ast }$ and
(d) $B^{+}_{c}\pi\rightarrow D^{\ast }B^{\ast }$ respectively, as a function of temperature.} \label{Fig.5}
\end{figure}%
\begin{figure}[H]
\begin{center}
\includegraphics[angle=0,width=.89\textwidth]{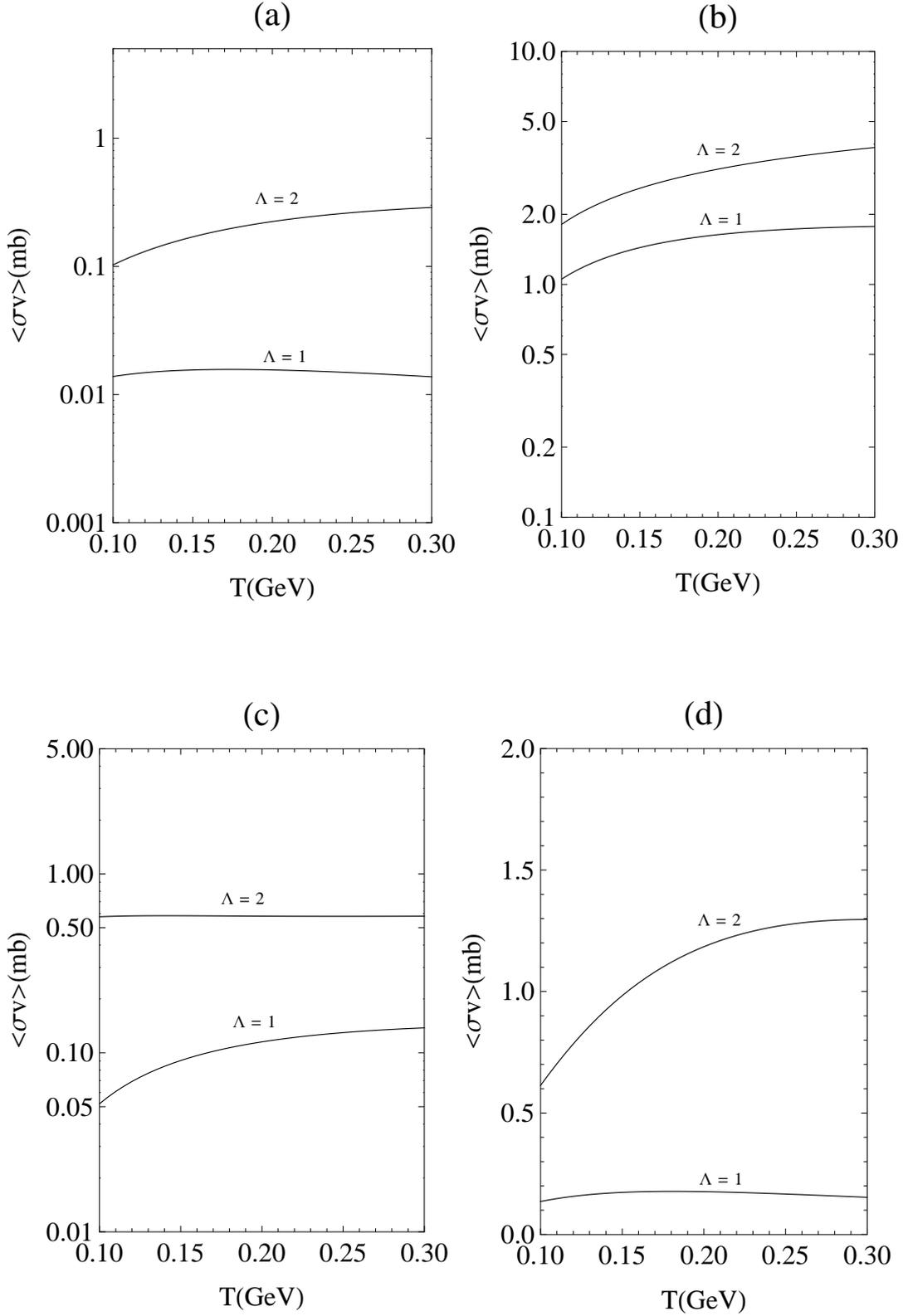}
\end{center}%
\caption{Thermal average $B_{c}$ dissociation process  of
(a) $B_{c}^{+}\rho\rightarrow DB$,
(b) $B^{+}_{c}\rho\rightarrow D^{\ast}B$,
(c) $B^{+}_{c}\rho\rightarrow D B^{\ast }$ and
(d) $B^{+}_{c}\rho\rightarrow D^{\ast }B^{\ast }$ respectively, as a function of temperature.}\label{Fig.6}
\end{figure}%

\newpage
\noindent Thus $\Psi$ can be expressed as

\begin{equation}
\Psi =\underset{a,b,c}{\sum }\langle\sigma_{bc\rightarrow B_{c}a}v_{bc}\rangle n_{b}n_{c}-\underset{a,b,c}{\sum }\langle \sigma _{B_{c}a
\rightarrow bc}v_{B_{c}a}\rangle n_{B_{c}}n_{a},
\end{equation}
where $\langle {\large \sigma } _{a B_{c}\rightarrow bc}{\large v}_{B_{c}a} \rangle$ and $\langle {\large \sigma } _{bc\rightarrow B_{c}a} {\large v}_{bc}\rangle$ represent the thermal average cross-section of $B_{c}$ dissociation with the co-moving particle $a$ and $B_{c}$ production through corresponding reverse processes, respectively. Imposing the simplifying assumption that the comovers almost remain in chemical equilibrium throughout the course of their interaction with $B_{c}$ mesons. This implies that the densities of $a$, $b$, and $c$ particles are supposed to have same equilibrium value at any temperature.
\begin{equation}
n_{a,b,c}\approx n_{a,b,c}^{eq}
\end{equation}

\noindent It means that the production rate of different particles except $B_{c}$ does not change throughout the interaction with comovers. In chemical equilibrium the principle of detailed balance holds and hence the rate of production of $B_{c}$ mesons is equal to its rate of absorption.
\begin{equation}
\langle{\large \sigma }_{B_{c}a\rightarrow bc} {\large v}%
_{B_{c}a}\rangle n_{B_{c}}^{{}eq}n_{a}^{eq}= \langle{\large \sigma}_{bc\rightarrow B_{c}a} {\large v}_{bc}\rangle n_{b}^{eq}n_{c}^{eq},
\end{equation}

\noindent where $n_{B_{c}}^{eq}$ is the initial value of $n_{B_{c}}$. The equilibrium density $n^{eq}$ of a hadron is given as
\begin{equation}
n^{eq}=\frac{dm^{2}T}{2\pi^{2}}K_{2}(m/T),
\end{equation}%
where $K_{2}$ is the modified Bessel function of second kind and second order, $m$ is the mass of the hadron,
and $d$ stands for the degeneracy factor (spin and isospin) of the hadron \cite{alvarez2002} and is given by
\begin{equation}
d=(2S+1)(2I+1).
\end{equation}
Here $S$ represents the spin and $I$ the isospin of the particle \cite{liu2006,alvarez2002}. Substituting Eqs. (22)-(24) in Eq. (\ref{rate eq0}), we get
\begin{equation}
\partial _{\mu } (n_{B_{c}}u^{\mu})=\underset{a,b,c}{\sum}\langle\sigma_{B_{c}a\rightarrow bc}v_{B_{c}a}\rangle(n_{B_{c}}^{eq}-n_{B_{c}})n_{a}^{eq}. \qquad
\end{equation}
\noindent Following the hydrodynamic model used in Ref. \cite{liu2006} in order to investigate the time evolution of the transverse radius of the fireball. In RHIC the particles are distributed almost uniformly in the central rapidity region. We use cylindrical coordinates $(\tau ,\eta,r,\phi )$ due to cylindrically symmetric geometry of collision. Here  $\tau$, $\eta$, $r$, and $\phi $ represent longitudinal proper time, space-time rapidity, transverse radius, and polar angle respectively \cite{liu2006}. The proper time $\tau$ and rapidity are defined as
\begin{equation}
\tau=(t^{2}-z^{2})^{\frac{1}{2}}, \ \  \eta=\frac{1}{2}\ln\frac{t+z}{t-z}.
\end{equation}
The density $n_{B_{c}}(\tau, \eta,r,\phi )$ remains constant in the $\phi -r$ plane due to cylindrical symmetry. The assumption of radial transverse expansion implies that $u^{\phi}=u^{\eta}=0$ \cite{liu2006}. Further assuming that in the transverse plane density of distribution is uniform, i.e., $u^{r}$ is constant. Applying these assumptions and averaging over the radial coordinate \cite{biro1983,xia1988}, we get
\begin{equation}
\frac{1}{\tau R^{2}(\tau)}\frac{\partial}{\partial\tau}(\tau R^{2}(\tau)n_{_{B_{c}}}\left\langle u^{\tau}\right\rangle)= \underset{a;b;c}
{\sum}\langle\sigma_{B_{c}a\rightarrow bc}v_{B_{c}a}\rangle(n_{B_{c}}^{eq}-n_{B_{c}})n_{a}^{eq}. \label{rate eq}
\end{equation}
Here $R(\tau)$ is the transverse radius of the fire-ball \cite{liu2006} and $\langle u^{\tau}\rangle$ represents the averaged $\tau$ component of four velocity vector which is expressed as
\begin{equation}
\langle u^{\tau} \rangle=\frac{2}{R^{2}(\tau)}\underset{0}{\overset{R(\tau)}{\int}} dr r u^{\tau}(r).
\end{equation}

\begin{figure}[H]
\begin{center}
\includegraphics[angle=0,width=0.9\textwidth]{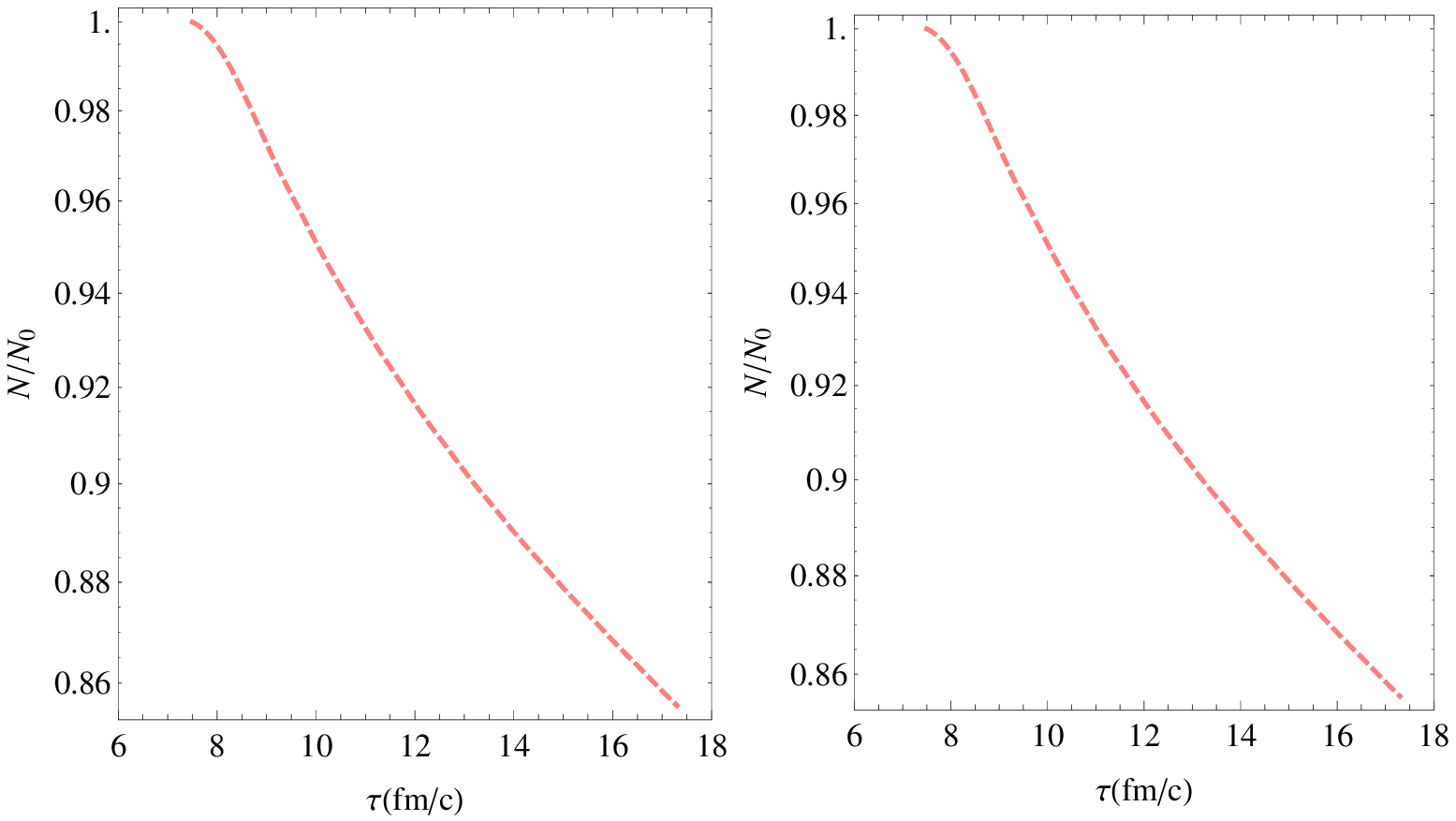}
\end{center}%
\caption{Time dependence of the ratio of the number of $B_{c}$ mesons for normal contribution from hadronization time $\tau_{H}=7.5$ fm/c to freeze out time $\tau_{F}=17.3$ fm/c. In the left panel for cutoff parameter $\Lambda = 1$\ GeV  and   in the right panel for $\Lambda = 2$ \; GeV.} \label{Fig.7}
\end{figure}

\begin{figure}[H]
\begin{center}
\includegraphics[angle=0,width=0.9\textwidth]{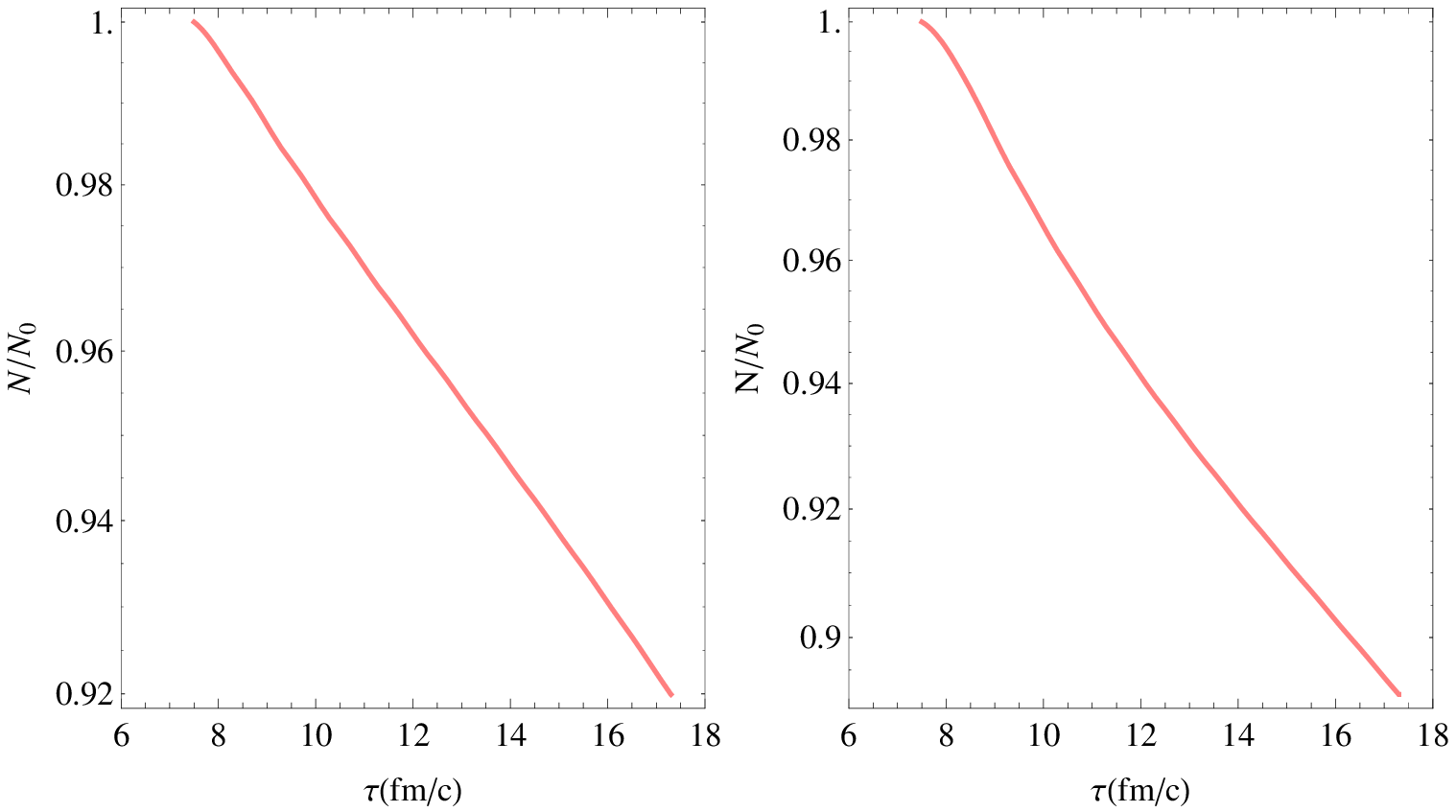}
\end{center}%
\caption{Time dependence of the ratio of the number of $B_{c}$ mesons for total contribution from hadronization time $\tau_{H}=7.5$ fm/c to freeze out time $\tau_{F}=17.3$ fm/c. In the left panel for cutoff parameter $\Lambda = 1$\ GeV and in the right panel for $\Lambda = 2$ \; GeV.} \label{Fig.7}
\end{figure}
 The expression of $\langle u^{\tau}\rangle$ in terms of $\beta_{r}$ (radial flow velocity of the hadronic matter) can be written as
\begin{equation}
 u^{\tau}=\frac{1}{\sqrt{1-\beta^{2}_{r}}}.
\end{equation}
$\beta_{r}$ is taken
\begin{equation}
\beta_{r}(\tau,r)=\frac{dR}{d\tau}\label{32}(\frac{r}{R})^{a}.
\end{equation}
\noindent Here $a$ is a constant and its value is taken to be 1 as in Refs. \cite{liu2006,alvarez2002}, Eq. (31) can be expressed as
\begin{equation}
\frac{dR}{d\tau}=\beta_{s}.
\label{betas}
\end{equation}
\noindent In Eq. (\ref{betas}) $\beta _{s}$ represents the transverse flow velocity of mid rapidity hadrons at RHIC. $\langle u^{\tau}\rangle$ in terms of $\beta _{s}$  can be written as
\begin{equation}
\langle u^{\tau }\rangle=\underset{0}{\overset{1}{\int}} dy\frac{1}{\sqrt{1-\beta_{s}^{2}y}}.
\end{equation}
\noindent The time evolution of transverse radius of fire ball is given as
\begin{equation}
R(\tau )=R_{H}+v_{H}(\tau -\tau _{H})+\frac{a}{2}(\tau -\tau _{H})^{2},
\end{equation}%

\noindent where $R_{H}\approx 9$ fm and $v_{H}\approx 0.4c$ \cite{liu2006}  represent transverse radius and transverse flow velocity of fire ball respectively at hadronization time $\tau_{H}=7.5 \;\textmd{fm/c}$ and $a=0.02 c^{2}$/fm. The values of these parameters are extracted from the measured values of transverse energy and transverse flow velocity of midrapidity hadrons in Au+Au collision at $\sqrt{s_{NN}}=200$ GeV. See Ref. \cite{liu2006} for details.
In Ref. \cite{liu2006} the relation between time and temperature of the hadronic matter is parameterized as
\begin{equation}
T(\tau )=T_{c}-(T_{H}-T_{F})(\frac{\tau -\tau _{H}}{\tau _{F}-\tau _{H}})^{0.8},
\end{equation}%

\noindent where $T_{c}$  is the critical temperature for QGP transition and $T_{H}$ is the hadronization temperature; we take $T_{c}=T_{H}=0.175$ GeV. $T_{F}=0.125$ GeV is
the freeze out temperature and $\tau_{F}\approx 17.3\;\textmd{fm/c}$ is the freeze out time.

\noindent By using thermal average cross-section described in the previous section and solving Eq. \ref{rate eq} numerically, the time dependence of of $B_{c}$ meson yield in hadronic matter at RHIC is calculated for $\Lambda = 1$ and $2$ GeV. The resultant time dependent yield normalized to equilibrium value is plotted in Fig. 7 from normal interaction and in Fig. 8 after including the anomalous interaction. Here the initial number of $B_{c}$ mesons is $N_{0}=\tau_{H} \pi R_{H}^{2}n_{B_{c}}^{eq}(T_{H})$ and the number of $B_{c}$ mesons at time $\tau$ is $N(\tau)=\tau \pi R^{2}(\tau)n_{B_{c}} (\tau)$. It can be seen from Fig. 7 that yield of  $B_{c}$ mesons is affected by almost $14\%$ for $\Lambda=1$ and 2 GeV. Fig. 8 shows that normalized yield of $B_{c}$ mesons slowly decreases with time in the hadronic matter. Total decrease which occurs is  $8\%$ for $\Lambda=1$ GeV and $11\%$ for $\Lambda=2$ GeV.

\section{Concluding remarks}
\noindent In this paper, we calculate cross sections for $B_{c}$ meson dissociation by $\pi$ and $\rho$ mesons using meson exchange model including anomalous couplings like PVV, PPPV and VVVP. Previously we have studied these processes without including these couplings. A knowledge of $B_{c}$ absorption cross sections by comovers (in this paper $\pi$ and $\rho$ mesons) is essential to extract information on properties of QGP at RHIC. As shown in Fig. $3$ the cross section for the process $ B^{+}_{c}+\pi\rightarrow D^{\ast }+B^{\ast }$ after including the anomalous terms is in the range of $3$ to $8$ mb, which is significantly enhanced as compared to our previous results for the same process which was $0.2$ to $2$ mb in Ref. \cite{lodhi2011}. From Fig. 4 it can be seen that the cross section for $B_{c}$ mesons by $\rho$ mesons is less than $1$ mb for all the processes away from the threshold except for the process $B^{+}_{c}+\rho\rightarrow D^{\ast }+B $ which is in the range of $ 2 $ to $9$ mb. To see the effects of these interactions on the $B_{c}$ meson yield at RHIC, we have studied its time evolution using the kinetic equation for the heavy ion collisions dynamics. The plot shows that the suppression caused by the interaction of $B_c$ mesons with comovers is almost $8\%$ and $11\%$ when $\Lambda=1$ and $2\;\textmd{GeV}$ respectively. These results show that although the effect of interaction with comovers is small but it is not negligible.


\bigskip
\begin{thebibliography}{99}
\bibitem{matsui1986} T. Matsui and H. Satz, Phys. Lett. B \textbf{178}, 416 (1986).

\bibitem{NA50} M. C. Abreu et. Al., NA50 Collaboration, Phys. Lett. B \textbf{450}, 456 (1999).

\bibitem{cassing1997} W. Cassing and C. M. Ko, Phys. Lett. B \textbf{396},39 (1996);\\
                      W. Cassing and E. L. Bratkovskaya, Nucl. Phys. A \textbf{623}, 570 (1997).

\bibitem{armesto1998} N. Armesto and A. Capella, Phys. Lett. B \textbf{430}, 23 (1998).

\bibitem{kahana1999} D. E. Kahana and S. H. Kahana, Phys. Rev. C \textbf{59}, 1651 (1999).

\bibitem{gale1999} C. Gale, S. Jeon and J. Kapusta, Phys. Lett. B \textbf{459}, 455 (1999).

\bibitem{spieles1999} C. Spieles, R. Vogt, L. Gerland, S. A. Bass, M. Bleicher, H. Stocker,\\
                      and, W. Greiner, Phys. Rev. C \textbf{60}, 054901 (1999).

\bibitem{sa1999} Ben-Hao Sa, An Tai, Hui Wang, and Geng-He Liu, Phys. Rev. C \textbf{59}, 2728 (1999).

\bibitem{kharzeev1994} D. Kharzeev and H. Satz, Phys. Lett. B \textbf{334}, 155 (1994).

\bibitem{sum-rule} D. Kharzeev, H. Satz, A. Syamtomov, and G. Zinovjev Phys. Lett. B \textbf{389}, 595 (1996).

\bibitem{quark models} C. Y. Wong, E. S. Swanson, and T. Barnes, Phys. Rev. C \textbf{62}, 045201 (2000);\\
                       M. A. Ivanov, J. G. Korner, and P. Santorelli, Phys. Rev. D \textbf{70}, 014005 (2004).

\bibitem{lin2000} Z. Lin and C. M. Ko, Phys. Rev. C \textbf{62}, 034903 (2000).

\bibitem{lin2001} Z. Lin and C. M . Ko, Phys. Lett. B \textbf{503}, 104-112 (2001).

\bibitem{haglin2000} Haglin, L. Kevin, Phys. Rev. C \textbf{61}, 031902 (2000).

\bibitem{liu2001} W. Liu, C. M. Ko, and Z. W. Lin, Phys. Rev. C \textbf{65}, 015203 (2001).

\bibitem{cms} The annual Quark Matter conference 2011, reported in CERN Bulletin Nos \textbf{21-22}, (2011).

\bibitem{vogt1997} R. Vogt, Phys. Rept. \textbf{310}, 197 (1997).

\bibitem{schro2000} Martin Schroedter, Robert L. Thews and Johann Rafelski, Phys. Rev. C \textbf{62},\\ 024905 (2000).

\bibitem{refelski2002} J. Letessier and J. Refelski, \emph{Hadrons and Quark-Gluon Plasma}\\
                       (cambridge University Press, UK), (2002).
\bibitem{lodhi2007} M. A. K Lodhi and Marshall, Rian., Nucl. Phys. A \textbf{790} 323c- 327c (2007).

\bibitem{lodhi2011} M. A. K Lodhi, Faisal Akram and Shaheen Irfan., Phys. Rev. C \textbf{84}, 03490.1 (2011)

\bibitem{faisal2011} Faisal Akram and M. A. K. Lodhi Phys. Rev.C \textbf{84}, 064912 (2011).

\bibitem{oh2001} Yongseok, Taesoo Song, and Su Houng Lee, Phys. Rev. C \textbf{65}, 034901 (2001).

\bibitem{azvedo2004}  R.S. Azevedo, and  M. Nielsen, Braz. J. Phys. \textbf{34}: 272-275 (2004).

\bibitem{faisal2012} Faisal Akram and M. A. K. Lodhi, Nucl. Phys. A \textbf{877}, 95-106 (2012).

\bibitem{pdg} Particle Data Group, D. E. Groom et al., Eur. Phys. J. C \textbf{15}, 1 (2000).

\bibitem{haglin2001} Haglin, L. Kevin, and Gale Charles, Phys. Rev. C \textbf{63}, 06520 (2001).

\bibitem{chan1997} L.-H. Chan, Phys. Rev. D \textbf{55}, 5362 (1997).

\bibitem{Pari1991} G. Pari, B. Schwesinger, and H. Walliser, Phys. Lett. B \textbf{255}, 1 (1991);\\
                   Y. Oh, D. P. Min, M. Rho, and N. N. Scoccola, Nucl. Phys. A \textbf{503} 534, 493 (1991).

\bibitem{liu2006} W. Liu a, C.M. Ko, L.W. Chen, Nuclear Physics A \textbf{765}, 401--425 (2006).

\bibitem{machleid1987} R. Machleid, K. Holinde and C. Elster, Phys. Rev. \textbf{149} , 1 (1987); R. Machleid, Adv. Nucl. Phys. \textbf{19}, 189
(1989); D. Lohse, J. W. Durso, K. Holinde, and J. Speth, Nucl. Phys. \textbf{A516}, 513 (1990).

\bibitem{yasui2009} S. Yasui and K. Sudoh, Phys. Rev. D \textbf{80}, 034008 (2009).

\bibitem{alvarez2002} L. Alvarez-Ruso and V. Koch, Phys. Rev. C \textbf{65}, 054901 (2002).

\bibitem{biro1983} T. Biro, H. W. Barz, B. Lukacs, and J. Zimanyi, Phys. Rev. C \textbf{27}, 2695 (1983).

\bibitem{xia1988} C. M. Ko and L.-H. Xia, Phys. Rev. C \textbf{38}, 179 (1988).

\end{thebibliography}
\end{document}